\documentclass[aps,prx,reprint,amsmath,amssymb,floatfix]{revtex4-2}

\usepackage{xcolor}
\usepackage{soul}
\usepackage[utf8]{inputenc}
\usepackage{physics}
\usepackage{amsfonts}
\usepackage{dcolumn}
\usepackage{bm}
\usepackage{wrapfig}
\usepackage[normalem]{ulem}
\usepackage[colorlinks=true, hyperindex, breaklinks, linkcolor=blue, urlcolor=blue, citecolor=blue]{hyperref}
\usepackage{lineno}
\usepackage[T1]{fontenc}

\usepackage{graphicx}
\usepackage{amssymb, amsthm, amsmath}
\usepackage{accents}
\usepackage{mathptmx} 
\usepackage{textcomp}
\usepackage{enumitem}

\usepackage{cleveref}
\crefname{equation}{Eq.}{Eqs.}
\Crefname{equation}{Eq.}{Eqs.}
\crefname{table}{Table}{Tables}
\Crefname{table}{Table}{Tables}
\crefname{figure}{Figure}{Figs.}
\crefname{appendix}{Appendix}{Appendices}
\Crefname{figure}{Fig.}{Figs.}
\crefname{section}{Section}{sections}
\Crefname{section}{Section}{Sections}
\crefname{subsection}{Section}{Sections}
\Crefname{subsection}{Section}{Sections}
\Crefname{appendix}{Appendix}{Appendices}

\usepackage[caption=false]{subfig}

\newcommand{\phantomsubfloat}[1]{
	{
		\captionsetup[subfigure]{labelformat=empty}
		\subfloat[][]{#1}
	}%
}

\newcommand{\pr}[2]{p\big(#1\,|#2\big)}

\DeclareMathOperator*{\argmax}{arg\,max}

\newcommand{\Gq}{\Gamma}
\newcommand{\Gup}{\Gq_\uparrow}
\newcommand{\Gphi}{\Gc_\phi}
\newcommand{\Gc}{\Gamma^\mathrm{c}}
\newcommand{\Gi}[1]{\Gamma^\mathrm{c}_{\phi,#1}}
\newcommand{\Gphifb}{\Gamma_{\phi,\mathrm{fb}}^\mathrm{c}}
\newcommand{\Tphi}{T_{\phi}^\mathrm{c}}
\newcommand{\Tphifb}{T_{\phi,\mathrm{fb}}^\mathrm{c}}
\newcommand{\Tphips}{T_{\phi,\mathrm{ps}}^\mathrm{c}}
\newcommand{\vunit}[3]{#1\,$\pm$\,#2\,#3}
\newcommand{\mus}{\textmu s}

\newcommand{\Tq}{T_1}
\newcommand{\tm}{t_\mathrm{m}}

\begin{document}

\title{\fontsize{11.25}{14}\selectfont Recovering Quantum Coherence of a Cavity Qubit Coupled \\ to a Noisy Ancilla through Real-Time Feedback}

\author{Uri Goldblatt}
\author{Nitzan Kahn}
\author{Sergey Hazanov}
\author{Ofir Milul}
\author{Barkay Guttel}
\author{Lalit M. Joshi}
\author{Daniel Chausovsky}
\author{Fabien Lafont}
\author{Serge Rosenblum}
\affiliation{
\vspace{3pt}Department of Condensed Matter Physics, Weizmann Institute of Science, Rehovot, Israel
}


\begin{abstract}
\noindent Decoherence in qubits, caused by their interaction with a noisy environment, poses a significant challenge to the development of reliable quantum processors. A prominent source of errors arises from noise in coupled ancillas, which can quickly spread to qubits. By actively monitoring these noisy ancillas, it is possible to not only identify qubit decoherence events but also to correct these errors in real time. This approach is particularly beneficial for bosonic qubits, where the interaction with ancillas is a dominant source of decoherence. 
In this work, we uncover the intricate dynamics of decoherence in a superconducting cavity qubit due to its interaction with a noisy transmon ancilla. By tracking the noisy ancilla trajectory and using real-time feedback, we successfully recover the lost coherence of the cavity qubit, achieving a fivefold increase in its pure dephasing time. Additionally, by detecting ancilla errors and converting them into erasures, we improve the pure dephasing time by more than an order of magnitude. These advances are essential for realizing long-lived cavity qubits with high-fidelity gates, and they pave the way for more efficient bosonic quantum error-correction codes.
\end{abstract}

\maketitle

Achieving long lifetimes of quantum information is crucial for reducing errors in quantum processors. This goal has driven significant advances in the coherence of superconducting quantum circuits 
\cite{Place2021NewMilliseconds,Wang2022TowardsMilliseconds,Manucharyan2009Fluxonium:Offsets,Grunhaupt2019GranularCircuits,somoroff2023millisecond}, which are particularly sensitive to noise. Despite these ongoing improvements, interaction with the noisy environment calls for error-mitigation strategies, such as quantum error correction (QEC) \cite{Krinner2022RealizingCode,Acharya2023SuppressingQubit,Zhao2022RealizationQubits,acharya2024quantum}.
This approach involves redundantly encoding information in entangled states of multiple physical qubits. Errors are detected through repeated measurement of the redundant degrees of freedom, and they are corrected using active feedback. 
While QEC can be effective regardless of the error's origin, it often requires significant hardware overhead.

A promising approach to reducing the overhead associated with QEC is to implement an underlying layer of error correction operating at the level of individual physical qubits \cite{Ofek2016ExtendingCircuits,kapit2016hardware}. This layer not only reduces error rates but may also provide information that can be leveraged by the QEC code. For instance, erasure qubits \cite{Tsunoda2023Error-DetectableAncilla,teoh2023dual,koottandavida2023erasure, chou2023demonstrating,levine2024demonstrating} allow for pinpointing the location of errors, thereby improving the threshold and effective distance of the QEC code \cite{Stace2009ThresholdsLoss,Kubica2022ErasureCircuits,wu2022erasure}. Alternatively,  suppressing a specific type of error \cite{Lescanne2020ExponentialOscillator,grimm2020kerr} can enable more efficient biased-noise codes to focus primarily on the most prevalent errors \cite{tuckett2018ultrahigh,xu2023tailored,ruiz2024ldpc}.

\begin{figure}[b!]
    \vspace{-15pt}    
    \centering    
    \includegraphics[scale=0.70]{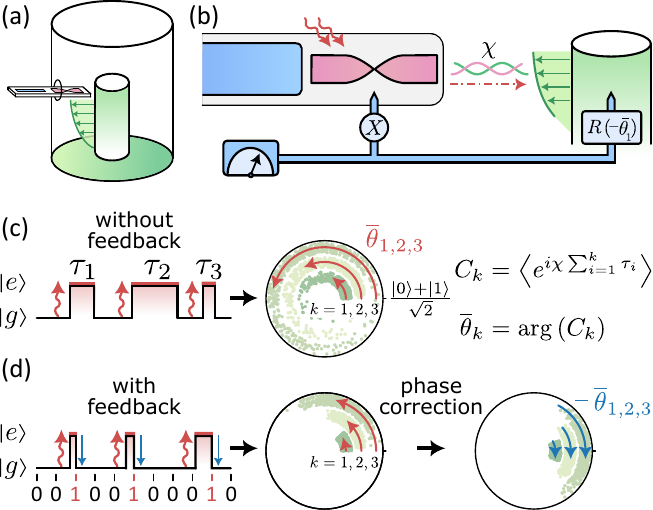}
    \phantomsubfloat{\label{fig:scheme a}}
	\phantomsubfloat{\label{fig:scheme b}}
	\phantomsubfloat{\label{fig:scheme c}}
	\phantomsubfloat{\label{fig:scheme d}}
    \vspace{-20pt}    
    \caption{\label{fig:scheme}\hspace{-1pt}\textbf{Coherence recovery in an ancilla-coupled cavity qubit.} \textbf{a}, Depiction of the experimental setup. A cavity qubit (green arrows) interacts dispersively with a chip-based transmon ancilla (magenta), which is monitored by a readout resonator (blue rectangle). \textbf{b}, Illustration of the coherence recovery protocol. Upon detecting an ancilla excitation (red wavy arrows), an ancilla reset pulse ($X$) and a cavity phase correction $R(-\bar{\theta}_1)$ are applied to minimize error propagation to the cavity (red dashed arrow). \textbf{c}, Illustration of the cavity qubit decoherence mechanism. Without feedback, each thermal ancilla excitation induces a cavity phase shift $\chi \tau_i$, with $\tau_i$ the exponentially distributed excitation duration and $\chi$ the dispersive interaction rate. As a result, the cavity state evolves from  $\frac{1}{\scriptscriptstyle{\sqrt{2}}}(\ket{0}\mkern-1mu+\mkern-1mu \ket{1})$ to $\frac{1}{\scriptscriptstyle{\sqrt{2}}}(\ket{0}\mkern-1mu+\mkern-1mu e^{i\theta_k}\ket{1})$, with $\theta_k=\chi\scriptstyle\sum_{i=1}^k\textstyle\tau_i$, representing the phase accumulated due to $k$ excitations. The distribution of $\theta_k$  is illustrated by individual realizations (green dots) on the equatorial plane of the cavity qubit's Bloch sphere. The uncertainty in $\theta_k$ affects the cavity coherence $C_k$, defined by the ensemble average of $e^{i\theta_k}$. Moreover, the coherence phase $ \bar\theta_k \equiv \arg(C_k)$ (red arrows) varies with the excitation count $k$, further contributing to dephasing. \textbf{d} Feedback protocol which uses ancilla reset pulses to limit excitation durations to the time between measurements, and a trajectory-dependent cavity phase correction $-\bar{\theta}_k=-k\bar{\theta}_1$. Both approaches minimize the cavity qubit's phase uncertainty.}
\end{figure}

The physical layer of error correction is not restricted to addressing qubit errors, but can also encompass errors occurring in coupled ancillary circuits. This is particularly relevant in superconducting bosonic systems \cite{Ma2021QuantumCircuits,Joshi2021QuantumQED}, such as three-dimensional cavities \cite{Romanenko2020Three-DimensionalS,Heidler2021Non-MarkovianMilliseconds,Milul2023SuperconductingTime,Chakram2021SeamlessElectrodynamics}, where ancillas such as transmons operating the bosonic qubits (\cref{fig:scheme a,fig:scheme b}) are a dominant source of decoherence \cite{Reagor2016QuantumQED, Ofek2016ExtendingCircuits,Campagne-Ibarcq2020QuantumOscillator,Sivak2023Real-timeBreak-even,Hu2019QuantumQubit,Ni2023BeatingQubit,pietikainen2024strategies}. While dispersive coupling to a transmon ancilla allows for universal control of the cavity qubit \cite{Heeres2017ImplementingOscillator,Krastanov2015UniversalQubit,Eickbusch2022FastQubit, Diringer2023ConditionalQubit}, it also causes thermal ancilla excitations to spread to the cavity in the form of frequency shifts. 
The uncertain durations of the excitations result in random phase accumulations of the cavity state, causing decoherence (\cref{fig:scheme c}). This shot-noise dephasing process presents a bottleneck to extending cavity qubit coherence times \cite{Reagor2016QuantumQED,Chakram2021SeamlessElectrodynamics,Milul2023SuperconductingTime} and to increasing the gain of bosonic quantum error-correction schemes \cite{Cai2021BosonicCircuits,Ofek2016ExtendingCircuits,Campagne-Ibarcq2020QuantumOscillator,Sivak2023Real-timeBreak-even,Hu2019QuantumQubit,Ni2023BeatingQubit}. In previous work  \cite{Rosenblum2018Fault-tolerantError.}, passive methods were used to decouple \cite{Zhang2017SuppressionQubit,Yan2016TheReproducibility} the transmon ancilla from the cavity qubit, thereby suppressing shot-noise dephasing. However, the absence of real-time feedback limited the achievable enhancements in cavity qubit coherence times.

In this work, we implement the \textit{in situ} correction of ancilla errors propagating to an idling cavity qubit by repeatedly measuring the ancilla and applying real-time feedback (\cref{fig:scheme d}). After every measurement, we reset the ancilla to its ground state \cite{Riste2012FeedbackMeasurement}, reducing the uncertainty in the duration of the excitations. In addition, we use the measurement outcomes to predict the average phase accumulated by the cavity qubit, allowing us to recover the encoded quantum information. 
Using this coherence recovery protocol, we show a fivefold increase in the pure dephasing time of the cavity qubit.  Moreover, the repeated measurements can be used to flag ancilla excitations, converting them to cavity qubit erasure errors \cite{Tsunoda2023Error-DetectableAncilla,teoh2023dual,koottandavida2023erasure, chou2023demonstrating,levine2024demonstrating}. Postselecting on no erasures, we observe an extension of the cavity pure dephasing time beyond 100 ms. Finally, we explore the protocol's performance for different measurement intervals and ancilla excitation rates.
Besides its application in quantum memories \cite{pietikainen2024strategies}, our approach also holds promise for quantum sensing \cite{Niroula2023QuantumQubits} and for realizing high-fidelity gates on cavity qubits \cite{Tsunoda2023Error-DetectableAncilla,teoh2023dual,reinhold2020error}.

We demonstrate our protocol using a three-dimensional aluminum stub cavity \cite{Reagor2016QuantumQED} whose fundamental mode has a single-photon lifetime of $T_1^\text{c} =$ \vunit{1.57}{0.02}{ms}. The transmon qubit, serving as an ancilla, has a lifetime $T_1 =$ \vunit{67.0}{0.3}{\mus}, a coherence time  $T_2 =$ \vunit{30.0}{0.2}{\mus}, and a thermal excited state population $\bar{n}_\mathrm{th}=$ \vunit{0.80}{0.05}{\%}. A readout resonator is dispersively coupled to the ancilla and connected to a measurement chain with quantum-limited amplification. This setup enables single-shot measurement of the ancilla state with a discrimination fidelity of 99.87\% (see \cref{sec:measurement_errors}). The transmon ancilla allows for universal control of the cavity qubit through their dispersive interaction, characterized by the Hamiltonian $\hat{H}/\hbar=-\chi\hat{n}\ketbra{e}{e}$. Here, $\chi/2\pi=73\,$kHz represents the dispersive interaction rate, $\ket{e}$ is the ancilla's first excited state, and $\hat{n}$ is the cavity photon number operator. This interaction results in a cavity phase-space rotation $e^{i\chi\tau\hat n }$ at an angle proportional to the total time $\tau$ the ancilla spends in the excited state (which we assume to be classical). Shot noise in the ancilla state due to its finite temperature therefore causes the cavity state to decohere.

Before describing our coherence recovery protocol in detail, we first characterize the baseline coherence time of our cavity qubit using a standard Ramsey-type experiment. We initialize the ancilla in the state $\frac{1}{\sqrt{2}}(\ket{g}+\ket{e})$, which we then map to the cavity using a sideband drive \cite{Pechal2014Microwave-controlledElectrodynamics,Rosenblum2018ACavities}. This procedure encodes a superposition of vacuum and the single-photon Fock state $\frac{1}{\sqrt{2}}(\ket{0}+\ket{1})$ in the cavity. Following varying idle periods, we decode the single-photon qubit by applying the sideband drive a second time. This method maps the single-photon qubit back to the ancilla, where its remaining coherence can be measured.
We observe a baseline cavity coherence time of $T_2^{\text{c}} = $ \vunit{2.24}{0.04}{ms}, enabling us to extract a pure dephasing time  $T_\phi^\mathrm{c}=(1/T_2^\mathrm{c}-1/2T_1^\text{c})^{-1} = $ \vunit{7.7}{0.3}{ms}. The pure dephasing rate expected from thermal ancilla excitations is approximately \cite{Wang2019CavityQubits}
\begin{equation}
\label{eq:gamma_phi_cav_no_fb}
   \Gphi = \Gup \frac{\chi^2}{\chi^2+\Gq^2}\approx \Gup,
\end{equation}
 with  $\Gq\equiv 1/\Tq$ the ancilla decay rate. The rate of thermal ancilla excitations is $\Gup\approx \bar n_\mathrm{th}\Gq=(8.4\pm 0.2\,\text{ms})^{-1}$, indicating that ancilla shot noise is the dominant cause of cavity  pure dephasing.

\begin{figure}[b!]
    \vspace{-18pt}
    \centering
    \includegraphics[scale=0.37]{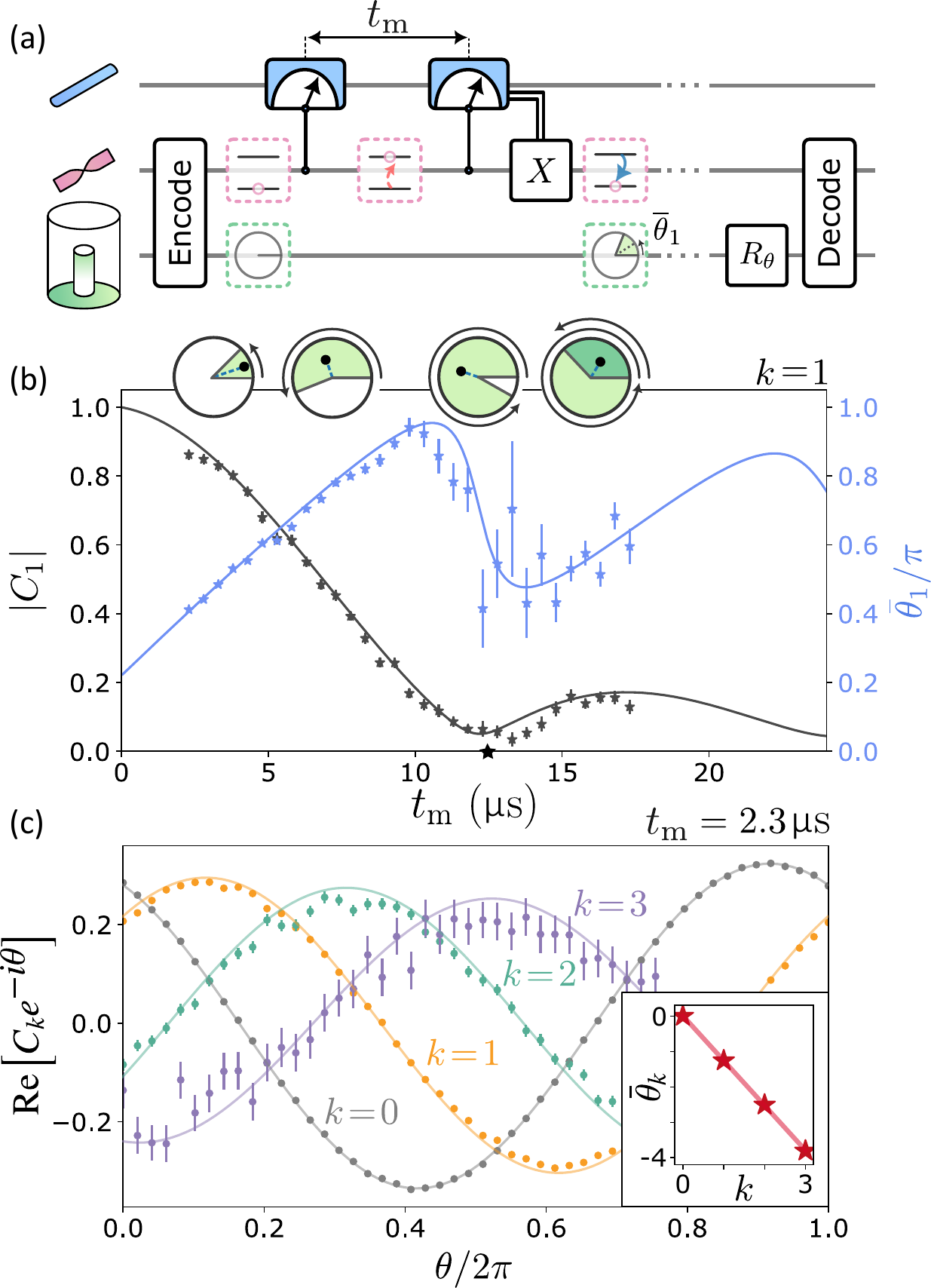}
    \vspace{-22pt}
    \phantomsubfloat{\label{fig:cavity_decoherence a}}
	\phantomsubfloat{\label{fig:cavity_decoherence b}}
	\phantomsubfloat{\label{fig:cavity_decoherence c}}
    \caption{\label{fig:cavity_decoherence}\hspace{-1pt} \textbf{Characterizing cavity decoherence with repeated ancilla measurements.} 
    \textbf{a}, Characterization circuit, which first encodes the cavity-ancilla system in $\frac{1}{\sqrt{2}}(\ket{0}+\ket{1})\otimes\ket{g}$. The ancilla is repeatedly measured and conditionally reset to $\ket{g}$ at intervals of $\tm$. The cavity state then undergoes a phase rotation $R_\theta$ and is subsequently decoded. Solid boxes represent operations, while dashed boxes show sample state trajectories. 
    \textbf{b}, Amplitude $|C_1|$ (black markers) and phase $\bar\theta_1$ (blue markers) of the extracted cavity state coherence as a function of $\tm$, conditioned on a single detected ancilla excitation ($k=1$). The coherence amplitude and phase are displayed relative to their values for $k=0$. The theoretical model (solid lines) is similar to \cref{eq:coherence_abs_and_phase} but accounts for ancilla decay (see \cref{sec:coherence}). From this model, we extract  $\chi/2\pi=82.1\,$ kHz and $\theta_0=0.22\pi$ (the experimental parameters in this figure differ from the rest of the article). The measurement interval corresponding to $\tm = 2\pi/\chi$ is indicated by a black star. The insets illustrate the equatorial plane of the cavity qubit’s Bloch sphere with the phase distribution in green and the coherence indicated by a black dot.     
    \textbf{c}, Cavity coherence postselected on different excitation numbers $k$ at $t=0.8\,$ms. The cavity state is measured along the basis $\frac{1}{\sqrt{2}}(\ket{0}\pm e^{i\theta}\ket{1})$ with varying $\theta$ by applying a rotation $R_\theta$ before the decoding. For $k=0$, coherence loss is mainly due to photon loss, while every added excitation reduces the coherence by a factor $|C_1|$ and shifts it by a phase $\bar{\theta}_1$. The phases $\bar{\theta}_k$ obtained from sinusoidal fits (solid lines) are plotted in the inset as a function of $k$, showing the expected linear dependence (red line).
    }
\end{figure}

As a first step towards addressing shot-noise dephasing, we monitor the ancilla in real time through repeated measurements \cite{Murch2013ObservingBit}. Upon detecting an excitation, we reset the ancilla to its ground state (\cref{fig:scheme d}). This procedure limits the duration of excitations to the time between measurements $\tm$, thereby reducing the cavity dephasing rate. 

To analyze the effect of the measurement interval $\tm$ on cavity decoherence, we focus on trajectories with a single detected excitation (\cref{fig:cavity_decoherence a}). This excitation induces a cavity phase shift $\theta_1=\chi \tau_1$, where the excitation duration $\tau_1$ is uniformly distributed between $[0,\tm]$, neglecting ancilla decay. The uncertainty in $\tau_1$ affects the cavity coherence, which is defined as the off-diagonal element of the cavity qubit's density matrix in the $Z$ basis \cite{baumgratz2014quantifying}. Specifically, a single excitation results in a cavity coherence $C_1=\langle e^{i\chi\tau_1}\rangle$, which corresponds to the characteristic function of $\tau_1$ (see \cref{sec:coherence}). The decoherence caused by this single excitation can be detected by measuring the cavity along the basis $\frac{1}{\sqrt{2}}(\ket{0}\pm e^{i\theta}\ket{1})$ with varying $\theta$. \cref{fig:cavity_decoherence b} shows the measured amplitude $|C_1|$ and phase $\bar\theta_1\equiv\arg(C_1)$ of the cavity coherence as a function of $\tm$. The results are in agreement with the theoretical prediction (see \cref{sec:coherence})

\begin{align}
\label{eq:coherence_abs_and_phase}
|C_1|&=\left|\text{sinc}\left(\frac{\chi \tm}{2}\right)\right|\nonumber\\
\bar\theta_1 &= \frac{\chi \tm} {2}\mathrm{mod}\,\pi+\theta_0,
\end{align}
where we assume ancilla decay is negligible, i.e.,  $\Gq\ll\chi,\tm^{-1}$ . The sinc profile arises because it represents the characteristic function of the uniform distribution, corresponding to the Fourier transform of a rectangular function. Similarly, $\frac{\chi \tm}{2}\mathrm{mod}\,\pi$ corresponds to the mean of the uniform distribution. The offset phase $\theta_0$ is a phenomenological parameter, which we attribute to the differential Stark shift from the measurement pulse when the ancilla is excited.

For short measurement intervals $\tm\ll\chi^{-1}$, the coherence can be approximated by $|C_1|\approx 1-\left(\chi \tm\right)^2/24$. In this limit, the excitation duration is too short to appreciably change the cavity phase, implying that frequent ancilla measurements can drastically mitigate cavity dephasing. Using a measurement interval of $\tm=2.6\,$\mus, the coherence loss due to a single excitation is limited to only $5.8\%$.
In contrast, for $\tm = 2\pi n/\chi $ with $n\in \mathbb{N}$, the phase distribution spans the entire unit circle, fully erasing the cavity coherence.

However, cavity pure dephasing arises not only due to the uncertain duration of individual ancilla excitations; the uncertainty in the number of such events also plays an important role (\cref{fig:scheme c}). The number of excitations $k$ observed up to time $t$ is approximately Poisson distributed with mean and variance given by $\Gup t$. The uncertainty in the resulting cavity phase $\theta_k= \chi\sum_{i=1}^k \tau_i$  contributes significantly to cavity dephasing. To evaluate the cavity coherence conditioned on $k$ excitations $C_k\equiv \langle e^{i \theta_k} \rangle$, we use the fact that for sufficiently low excitation rates $\Gup\ll \Gq$, the phases acquired from different excitations are mutually independent. Consequently, we obtain $C_k=(C_1)^k$. The observed coherence postselected on $k$ excitations (\cref{fig:cavity_decoherence c}) confirms that every added excitation shifts the coherence by a constant phase $\bar\theta_1$ and reduces its amplitude by a fixed factor $|C_1|$.

To minimize the decoherence arising from the uncertainty in the number of excitation events, we apply a phase correction $-\bar\theta_k=-k\bar\theta_1$ before decoding the cavity state (\cref{fig:scheme d}), where $k$ is the observed excitation count for each individual experimental sequence. Equivalently, we can apply a phase correction $-\bar{\theta}_1$ in real time each time an excitation is observed.

\begin{figure}[t!]
    \centering
    \includegraphics[scale=0.3]{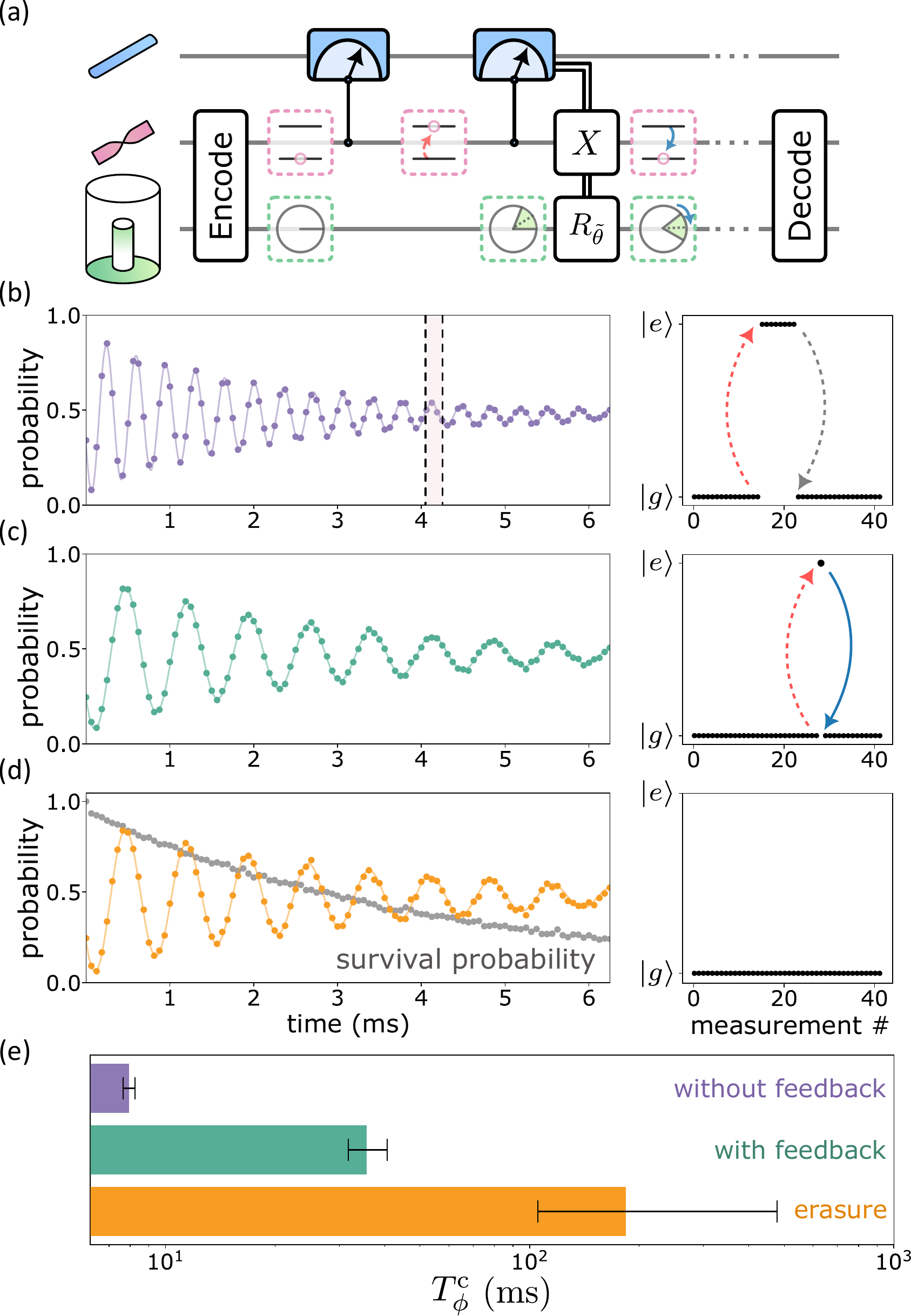}
    \phantomsubfloat{\label{fig:feedback a}}
	\phantomsubfloat{\label{fig:feedback b}}
	\phantomsubfloat{\label{fig:feedback c}}
    \phantomsubfloat{\label{fig:feedback d}}
    \phantomsubfloat{\label{fig:feedback e}}
    \vspace{-30pt}
    \caption{\label{fig:feedback}\hspace{-1pt} \textbf{Enhancing quantum coherence through ancilla monitoring and real-time feedback.} \textbf{a}, Circuit for the coherence recovery protocol, involving the application of an ancilla measurement followed by a conditional reset pulse $X$ and cavity phase correction $R_{\widetilde\theta}$ in a repetitive cycle with measurement interval $\tm=2.6$\,\mus. Solid boxes represent circuit operations, while dashed boxes show sample state trajectories.  In panels \textbf{b-d}, the left panels show the probability of measuring the ancilla in $\ket{e}$ after the decoding sequence, yielding cavity coherence decay curves. The right panels display representative ancilla trajectories taken from a short segment of the experiment (shaded area). Spontaneous excitation (relaxation) events are shown by red (gray) arrows, while the active reset operation is depicted by a blue arrow. We compare outcomes for three cases:  \textbf{b}, without measurement and feedback, \textbf{c}, with both measurement and feedback, choosing a correction phase $\widetilde\theta$ that maximizes the coherence, and \textbf{d}, removing erasure errors, i.e., keeping only experimental shots without observed ancilla excitations. The corresponding coherence times are $2.24\pm 0.04$ ms, $2.85\pm0.04$ ms, and $3.10\pm 0.06$ ms, respectively. Measuring the fraction of experimental shots that survive after postselection (gray markers) reveals an exponentially decaying survival probability with erasure rate $(4.43\,\mathrm{ms})^{-1}$. \textbf{e}, Comparison of pure dephasing times for the three cases, indicating an improvement from \vunit{7.7}{0.3}{ms} for the no-feedback case to \vunit{35}{4}{ms} with feedback. Postselecting on no erasures further enhances the pure dephasing time to $182^{+292}_{-78}$ ms.
    The quoted pure dephasing times are the medians of the most likely probability distributions consistent with the data in panels b-d, with error bars indicating the 68\% confidence interval.
    } 
    \vspace{-20pt}
\end{figure}

Having set up all necessary elements, we proceed to implement the coherence recovery protocol. This protocol involves the repetitive execution of three steps: high-fidelity measurement of the ancilla state, conditional reset of the ancilla, and conditional correction of the accumulated cavity phase (\cref{fig:feedback a}). 
The reset pulse reduces the uncertainty in the duration of each excitation and prevents further excitations to the second excited ancilla state $\ket{f}$. The phase correction removes the contribution from the uncertainty in the number of excitations, leaving the residual uncertainty from individual excitations as the only contribution to cavity dephasing.
 Since the rate of such events is $\Gup$, the coherence recovery protocol is predicted (see \cref{sec:coherence}) to produce a cavity pure dephasing rate
\begin{equation}
\label{eq:coherence_no_error_with_fb}
\Gphifb=\Gup\left(1-\left|\text{sinc}\left(\frac{\chi \tm}{2}\right)\right|\right),
\end{equation}
yielding an expected improvement factor of  $  \Gphi/\Gphifb\approx 17$ with respect to the idling case. 

An alternative strategy without the need for feedback could be considered, wherein ancilla excitations are autonomously dissipated. For example, a parametric drive can transfer the excitation to a lossy resonator, where the energy is quickly dissipated \cite{Magnard2018FastQubit}. Although this approach is similar to the application of conditional reset pulses, it is unable to correct for cavity phase accumulations. Consequently, the expected pure dephasing times are 4 times lower than those achievable with feedback (see \cref{sec:coherence}).

We evaluate the performance of the coherence recovery protocol by initializing the cavity to $\frac{1}{\sqrt{2}}(\ket{0}+\ket{1})$, applying the protocol for varying periods, and measuring the remaining coherence. Using our shortest measurement interval $\tm = 2.6$\,\mus, we find that the coherence time is extended from $T_2^{\text{c}} = 2.24\pm0.04\,\text{ms}$ to $2.85\pm0.04\,\text{ms}$ (\cref{fig:feedback b,fig:feedback c}). This finding corresponds to an enhancement of the pure dephasing time by a factor of $4.5$, from $\Tphi =$ \vunit{7.7}{0.3}{ms} to $\Tphifb =$ \vunit{35}{4}{ms} (\cref{fig:feedback e}). Though substantial, this improvement does not reach the expected factor of 17, which is primarily due to false-positive errors, where the ancilla is measured in $\ket{e}$, even though no excitation has occurred. These measurement errors trigger unnecessary reset pulses, which excite the ancilla. While most of these spurious excitations are corrected in the subsequent measurement round, we bias the measurement decision boundary to minimize their occurrence (see \cref{sec:measurement_errors}). We obtain a false-positive probability of $p_{e|g}=2.2\times 10^{-4}$, defined as the probability of measuring $\ket{e}$ when the ancilla is in $\ket{g}$. This finding corresponds to a spurious excitation rate of $\frac{p_{e|g}}{\tm}= (11.8\,\text{ms})^{-1}$.

In the context of quantum error-correcting codes, the ability to flag qubit errors can substantially improve error thresholds and code distances \cite{Stace2009ThresholdsLoss,Kubica2022ErasureCircuits,chou2023demonstrating,koottandavida2023erasure,levine2024demonstrating}. Converting ancilla excitations into such ``erasure errors'' of the cavity qubit can therefore be of great benefit, provided that the residual error rate after postselecting on no erasures is substantially smaller than the erasure rate. 
Excluding all experimental runs in which ancilla excitations were observed, we obtain a coherence time of $3.10\pm0.06\,\text{ms}$, close to the 2$T_1^c$ limit imposed by single-photon loss (\cref{fig:feedback d}). The corresponding pure dephasing time is $\Tphips = \, 182^{+292}_{-78}\,$ms (\cref{fig:feedback e}). We mostly attribute the remaining dephasing to shot-noise cavity dephasing induced
by the measurement pulses (se \cref{fig:supp_ancilla_error}). The observed erasure rate $(4.43\,\mathrm{ms})^{-1}$ is in agreement with the expected excitation rate of $\Gup+\frac{p_{e|g}}{\tm} = (4.7\,\mathrm{ms})^{-1}$. The erasure rate is, therefore, a factor of 41 larger than the pure dephasing rate after postselecting on no erasures. In the case of bosonic qubits with high photon lifetimes \cite{Romanenko2020Three-DimensionalS,Milul2023SuperconductingTime}, this method can provide the necessary hierarchy of error rates for its application as an erasure qubit.

\begin{figure}[b!]
    \vspace{-6pt}
    \centering
    \includegraphics[scale=0.35]{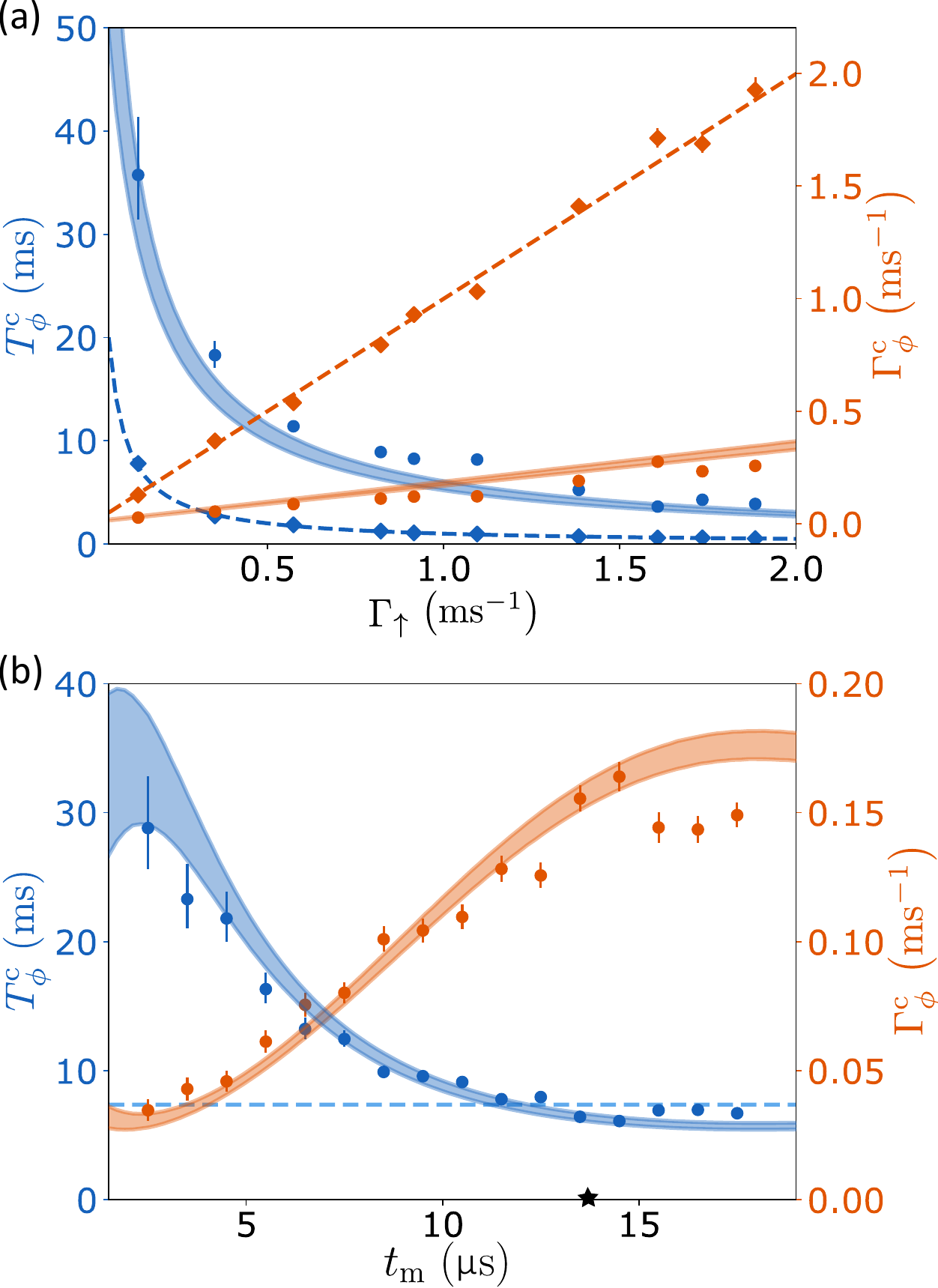}
    \phantomsubfloat{\label{fig:feedback_perf a}}
	\phantomsubfloat{\label{fig:feedback_perf b}}
    \vspace{-20pt}
    \caption{\label{fig:feedback_perf}\hspace{-1.2pt} \textbf{Performance of the coherence recovery protocol.} \textbf{a}, Cavity pure dephasing times $\Tphi$ (blue) and rates $\Gphi=1/\Tphi$ (orange) for various ancilla heating rates without measurement and feedback (diamond markers) and with measurement and feedback (circles) using $\tm=2.6$\,\mus. The theoretical models are shown with dashed and solid lines for the idling and feedback cases, respectively (see \cref{sec:measurement_errors}). The shading indicates the uncertainty in the experimental parameters. \textbf{b}, Cavity pure dephasing times (blue) and rates (orange) as a function of the measurement interval $\tm$. The pure dephasing time for an idling cavity qubit is shown by a dashed blue line. The black star denotes the measurement interval corresponding to $\tm=2\pi/\chi$.}
\end{figure}

Next, we examine the robustness of the coherence recovery protocol by characterizing its performance across a range of experimental parameters.
First, we characterize the pure dephasing rate as a function of the thermal excitation rate, both with and without applying the feedback protocol. We achieve this result by applying a weak incoherent ancilla drive with varying power, thereby inducing an effective ancilla heating mechanism. For sufficiently small thermal populations, we expect the pure dephasing rate to be linear \cite{Yan2016TheReproducibility} in the thermal excitation rate (see \cref{eq:gamma_phi_cav_no_fb,eq:coherence_no_error_with_fb}). In fact, for the idling case, we observe that the pure dephasing rate approximately equals the heating rate (\cref{fig:feedback_perf a}). This result is expected since, in the limit $\chi \gg \Gamma$, every ancilla excitation completely decoheres the cavity state (see \cref{eq:gamma_phi_cav_no_fb}). The data for the feedback case also match the expected linear dependence on the excitation rate.  
The ratio of the two slopes indicates a $7.3$-times improvement in pure dephasing times with the coherence recovery protocol. This improvement exceeds the one observed without inducing ancilla heating, as higher excitation rates reduce the significance of false positive errors and measurement-induced cavity dephasing.

Next, we investigate the performance of the protocol for varying measurement intervals $\tm$. For this experiment, we choose a correction phase of $\chi \tm/2$. While this phase is optimal for short measurement intervals, it deviates from the optimal phase for $\tm\gtrsim 2\pi/\chi$. The observed results agree with the expected behavior (\cref{fig:feedback_perf b}). Peak performance is achieved at the highest measurement rate. Further increasing the measurement rate would result in diminishing returns and eventually degrade the coherence, as measurement-induced cavity dephasing and false-positive errors start to dominate.  Conversely, for lower measurement rates, the residual uncertainty per excitation event grows, leading to increased dephasing rates. When the measurement rate is below $\chi/2\pi$, the phase correction deviates by more than $\pi/2$ from the optimal phase, resulting in a dephasing rate higher than that of the idling case.

In this work, we have demonstrated a significant enhancement in the pure dephasing time of cavity qubits by monitoring the coupled transmon ancilla and applying real-time feedback. The protocol effectively mitigates dephasing errors across a wide range of experimental parameters. The enhancement is even more pronounced if detected ancilla excitations are treated as erasure errors. One promising improvement is to decide in real time whether to treat a detected ancilla error as an erasure error, or whether to attempt to correct the cavity state. This method could reduce the erasure rate without significantly increasing computational errors. The results could be further improved by using a state estimation algorithm that takes into account the observation history, as opposed to a decision based only on the current measurement (see \cref{sec:markov}). This method could be used to tailor optimized correction phases for a wider range of identified events, such as measurement errors.

The current demonstration focuses on correcting dephasing errors while the cavity qubit is idling. However, our approach could also be applied during single-qubit or two-qubit gates \cite{Tsunoda2023Error-DetectableAncilla,teoh2023dual,reinhold2020error}. For instance, ancilla measurements can be applied when the ancilla is expected to be in the ground state during the gate operation, which allows for the detection and correction of ancilla decoherence events, thereby potentially enhancing gate fidelities.
Furthermore, although demonstrated with single-photon qubits, our method can also be applied to multiphoton qubits, potentially solving a major bottleneck in bosonic quantum error correction \cite{Ofek2016ExtendingCircuits,Sivak2023Real-timeBreak-even,Campagne-Ibarcq2020QuantumOscillator,Hu2019QuantumQubit,Ni2023BeatingQubit}.

Our method may serve as a physical layer of error correction within a larger quantum error-correction framework. By correcting pure dephasing errors, this layer could yield physical qubits with biased noise \cite{tuckett2018ultrahigh}, or, through error flagging, produce qubits suitable for erasure correction \cite{wu2022erasure}. Both outcomes can enhance the performance of higher-level error-correction layers. 

We anticipate that our method will provide the most significant benefits for qubits with long single-photon lifetimes, such as those recently demonstrated using niobium cavities \cite{Milul2023SuperconductingTime}. For these qubits, monitoring the transmon ancilla and actively correcting or erasing dephasing errors could lead to unprecedented quantum memory lifetimes, gate fidelities, and bosonic error-correction gains.
\\

\begin{acknowledgments}

We acknowledge valuable input from Vladimir Sivak, Alec Eickbusch, Ofer Zeitouni, and Barak Zackay. We acknowledge financial support from the Israel Science Foundation ISF Quantum Science and Technologies Grants 963/19 and 2022/20, and the European Research Council Starting Investigator Grant Q-CIRC 101040179. S.R. is the incumbent of the Rabbi Dr. Roger Herst Career Development Chair.
\end{acknowledgments}

\appendix

\let\oldsection\section
\renewcommand{\section}[1]{\oldsection{\MakeUppercase{#1}}}

\section{Device specification}
\begin{figure*}[b!]
    \centering
    \includegraphics[scale=0.35]{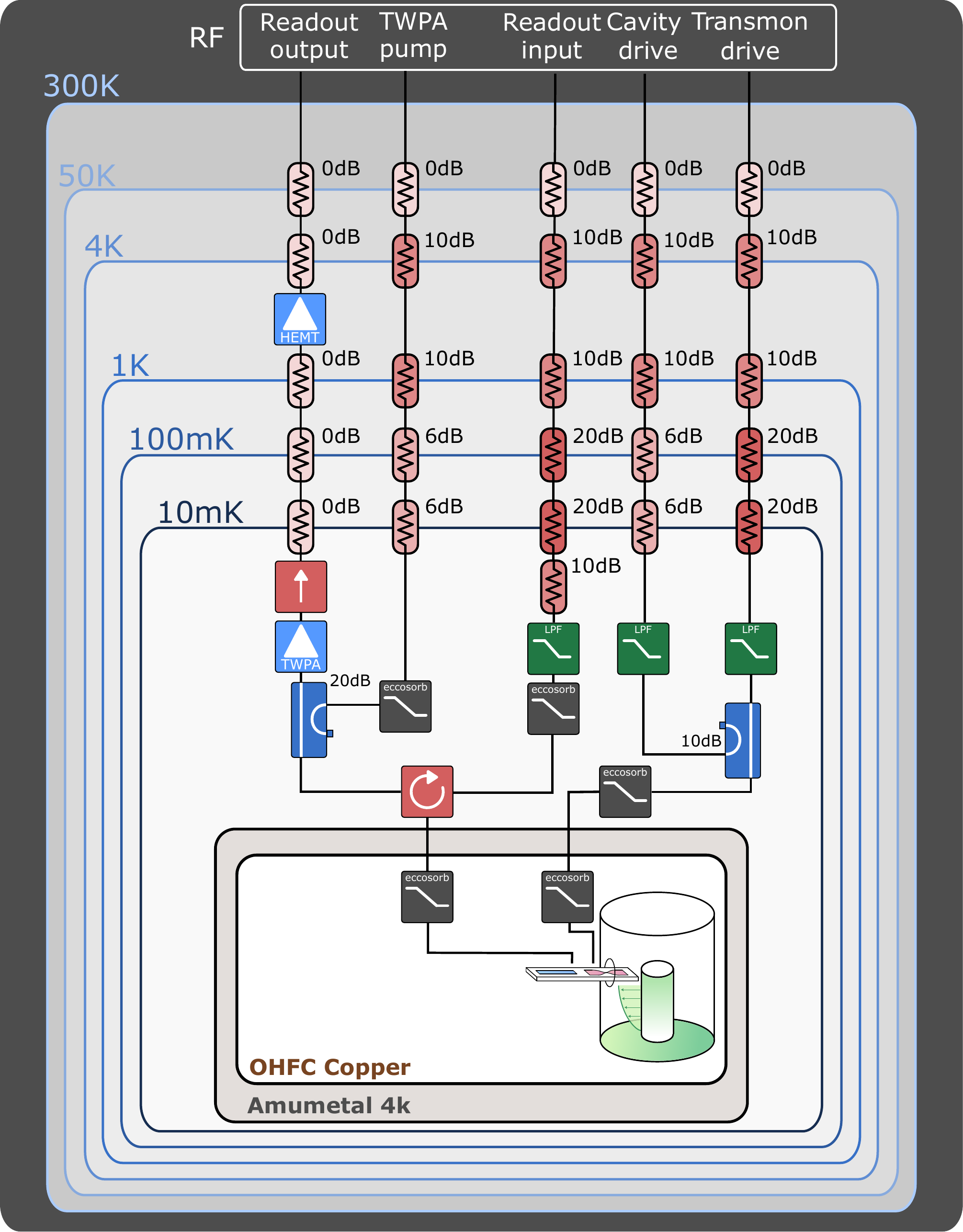}
    \caption{\label{fig:supp_device_specification} \textbf{Wiring diagram of the cryogenic microwave setup}. The experimental device is placed inside OFHC and A4K shields. The control signals are generated using Quantum Machines' OPX+ system and upconverted using IQ mixers. The signals are passed through a series of attenuators, low pass filters (LPF), and Eccosorb infrared filters. The output signal is amplified by a traveling-wave parametric amplifier (TWPA) from Silent Waves before passing through a double-junction isolator, enabling high-fidelity single-shot measurements.}
    
\end{figure*}

The experimental device is comprised of three elements: a cavity, a transmon ancilla, and a readout resonator. The cavity is made from 5N5 aluminum and has a coaxial stub geometry. The cavity was etched to remove contaminants and surface damage introduced by machining. The aluminum transmon ancilla and readout resonator are fabricated on a sapphire chip using electron beam evaporation. The chip is inserted into a coaxial waveguide connected to the cavity, and accessed through input and output couplers in the coaxial waveguide. The device is placed inside a thermalized oxygen-free high conductivity (OFHC) copper shield and encapsulated by an  Amumetal 4K (A4K) magnetic shield (see \cref{fig:supp_device_specification}).

\section{Device Hamiltonian and parameters}
\label{sec:hamiltonian_and_parameters}
We can express the free evolution of the system using the following Hamiltonian:
\begin{align*}
    \hat{H} = &\omega_r\hat{r}^\dagger\hat{r} + \omega_q\hat{q}^\dagger\hat{q} + \omega_c\hat{c}^\dagger\hat{c} 
    - \frac{K_r}{2}{\hat{r}^\dagger}^2\hat{r}^2
    - \frac{K_q}{2}{\hat{q}^\dagger}^2\hat{q}^2 \\
    &\, - \frac{K_c}{2}{\hat{c}^\dagger}^2\hat{c}^2
    - \chi\hat{c}^\dagger\hat{c}\hat{q}^\dagger\hat{q}
    - \chi_{\text{qr}}\hat{r}^\dagger\hat{r}\hat{q}^\dagger\hat{q}
    - \chi_{\text{cr}}\hat{r}^\dagger\hat{r}\hat{c}^\dagger\hat{c},
\end{align*}
where $\hat{r}, \hat{q}, \hat{c}$ are the annihilation operators of the readout, transmon ancilla, and cavity modes, respectively. The values of the parameters are listed in \cref{table:1}.

\setlength\extrarowheight{3pt}
\setlength{\tabcolsep}{8pt}
\begin{table*}[t!]
\centering
\begin{tabular}{||m{1.5cm} m{6cm} m{5.5cm} ||} 
 \hline
 \textbf{Parameter} & \textbf{Description} & \textbf{Value} \\ [0.5ex] 
 \hline\hline
 $\omega_c/2\pi$ & Cavity resonance frequency & 4.45 GHz \\ 
 $T_1^c$  & Cavity single-photon lifetime & \vunit{1.57}{0.02}{ms} \\ 
 $T_2^c$  & Cavity coherence time & \vunit{2.24}{0.04}{ms} \\ 
 $K_c/2\pi$ & Cavity anharmonicity & \vunit{14.3}{0.7}{Hz} $^{(*)}$ \\
 $\chi/2\pi$ & Cavity-transmon dispersive shift & \vunit{73.06}{0.06}{kHz} \\
 $\chi_{\text{cr}}/2\pi$ & Cavity-readout dispersive shift & \vunit{552}{18}{Hz} $^{(*)}$ \\
 \hline
 $\omega_\mathrm{q}/2\pi$ & Readout resonance frequency & $3.93\,\text{GHz}$ \\
 $K_\mathrm{q}/2\pi$ & Readout anharmonicity & $128\,\text{MHz}$  \\
 $\Tq$ & Readout lifetime & \vunit{67.0}{0.3}{\mus}\quad  (\vunit{31.5}{1.0}{\mus})\\
 $T_2$ & Readout coherence time & \vunit{29.97}{0.19}{\mus}\\
 $T_{2E}$ & Readout Hahn-echo coherence time & \vunit{68}{1}{\mus}  \\
 $\bar{n}_{\text{th}}$ & Readout average thermal population & \vunit{0.80}{0.05}{\%}\quad (\vunit{0.43}{0.04}{\%})  \\
 $\Gup$ & Readout excitation rate & \vunit{119}{5}{Hz} \quad(\vunit{134}{11}{Hz})  \\

 $\chi_{\text{qr}}/2\pi$ & Readout-readout dispersive shift & \vunit{1.44}{0.02}{MHz}  \\
 \hline
 $\omega_r/2\pi$ & Readout resonance frequency & $7.83\,\text{GHz}$ \\
$T_r$ & Readout lifetime & \vunit{607}{8}{ns} \\
$K_r/2\pi$ & Readout anharmonicity & \vunit{4.9}{0.2}{kHz} $^{(*)}$ \\
 \hline
\end{tabular}
\caption{System parameters and their respective values. Values in parentheses indicate parameters in the presence of repetitive measurements, extracted using a hidden Markov model (see \cref{sec:markov}). In addition, values obtained through simulation, rather than direct measurement, are indicated with an asterisk.}
\label{table:1}
\end{table*}

\section{Analysing cavity decoherence caused by a noisy ancilla}
\label{sec:coherence}
In this appendix, we examine the coherence of a cavity qubit initially prepared in $\frac{1}{\sqrt{2}}(\ket{0}+\ket{1})$ and dispersively coupled to a noisy transmon ancilla. We ignore cavity photon loss and focus only on the phase accumulation due to the ancilla excitations.

\subsection{Measuring the cavity coherence}
\vspace{-10pt}
\noindent The coherence of the cavity qubit is characterized by measurements along the basis $\ket{\psi_\theta^\pm}\equiv \frac{1}{\sqrt{2}}(\ket{0} \pm e^{i\theta}\ket{1})$ with varying $\theta$,  
which is achieved by adjusting the phase of the sideband drive in the decoding sequence and then measuring the ancilla in the computational basis.
The probability of observing  $\ket{\psi_{\theta}^+}$ is given by
\begin{align*}
    P_\theta &\equiv \Tr(\rho \ketbra{\psi_\theta^+})=\frac{1}{2}\Big(1 + \mathcal{C} e^{-i\theta}\Big) 
    \\&=\frac{1}{2}\Big(1+|\mathcal{C}| \cos (\bar\theta-\theta)\Big),
\end{align*}
where the cavity coherence $\mathcal{C}$ is defined by the off-diagonal element of the cavity qubit's density matrix  $\rho$  in the $Z$ basis \cite{baumgratz2014quantifying}.
The coherence amplitude $|\mathcal{C}|$ and phase $\bar\theta \equiv \mathrm{arg}(\mathcal{C})$ are determined by fitting the data for $2P_\theta-1=\Re{
    \mathcal{C} e^{-i\theta}}$ for varying $\theta$ to a sinusoidal curve (see \cref{fig:cavity_decoherence c}). This process yields
\vspace{-5pt}
\begin{align*}
    |\mathcal{C}| = \max_{\theta}\abs{2P_\theta - 1} ,\qquad 
    \bar\theta = \argmax_{\theta}\left(2P_\theta - 1\right).
\end{align*}
Similarly, the coherence conditioned on observing $k$ excitations, denoted by $C_k$, is obtained by postselecting the measured data  $P_{\theta}$ on observing $k$ excitations and fitting the resulting data.

\subsection{Cavity coherence in the presence of a noisy ancilla}
In a single experimental shot, the cavity evolves to $\ket{\psi_\theta} = \frac{1}{\sqrt{2}}(\ket{0}+e^{i\theta}\ket{1})$, where the total accumulated phase $\theta$ is a realization of the random variable $\Theta=\chi T$ with $T$ the total duration the ancilla spent in $\ket{e}$.
We can express the density matrix of the cavity qubit after tracing out the ancilla as an ensemble of different phase accumulations:
\begin{align*}
    \rho = \int \ket{\psi_{\chi\tau}}\bra{\psi_{\chi\tau}}\,f_T(\tau)\,d\tau,
\end{align*}
where $ f_T(\tau)$ is the probability distribution of $T$.
The cavity qubit coherence $\mathcal{C}$ is then given by the off-diagonal element of the cavity's density matrix 
\begin{align*}
\mathcal{C}\equiv\rho_{10}=\int  \braket{1}{\psi_{\chi\tau}}\braket{\psi_{\chi\tau}}{0} f_T(\tau) \, d\tau = \langle e^{i\chi T} \rangle,
\end{align*}

where $\langle g(T)\rangle$ represents the expected value of any function $g$ of the random variable $T$. 
The last equality suggests that in our scenario, we can interpret the coherence as the characteristic function of $T$. 

\subsection{Deriving the cavity coherence}
\label{sec:coherence_derivation}

 We consider the scenario where the ancilla is subjected to periodic measurements at intervals of $\tm$, followed by a reset to the ground state if an excitation is detected.
A key realization is that the random variables $T_i$ corresponding to the excited-state durations in measurement intervals indexed by $i$ are independent and identically distributed due to the reset operation. As such, we can write for the coherence at time $t$:
\begin{align}
    \mathcal{C}(t) &= \langle e^{i\chi\sum_{i=1}^{t/\tm} T_i} \rangle = \langle \prod_{i=1}^{t/\tm} e^{i\chi T_i} \rangle \nonumber \\
    &= \prod_{i=1}^{t/\tm} \langle e^{i\chi T_i} \rangle = C^{t/\tm} = e^{\frac{\ln (C)}{\tm}t},
\end{align}
where $C \equiv \langle e^{i\chi T_1} \rangle $ is the coherence after a single measurement interval.
If the decoherence during a single measurement interval is small, we can approximate the cavity pure dephasing rate by $\Gphifb = -\Re(\ln(C))/\tm\approx(1-\Re(C))/\tm$.

Next, we consider that within a single measurement period, several mutually exclusive events can occur, each with its own associated coherence $C^{(j)}$ and probability of occurrence $p_j$. We can therefore write $C=\sum_j {p_j C^{(j)}}$.
The dephasing rate is then 
\begin{align}
    \Gphifb &\approx \frac{1}{\tm} \Big(1- \sum_j {p_j \Re (C^{(j)})}\Big) \nonumber\\
    &= \frac{1}{\tm}\sum_j {p_j\left(1- \Re(C^{(j)})\right)} = \sum_j {\Gi{j}},
\end{align}
where $\Gi{j} \equiv \frac{1}{\tm} p_j\left(1- \Re (C^{(j)})\right)$ is the contribution of the $j$th dephasing mechanism to the total cavity pure dephasing rate. We now enumerate the possible events and derive their associated coherences and probabilities. In the current section, we assume a simplified model with only ancilla excitation and decay events. In the next section, we will include measurement errors and other imperfections. We assume $\Gup \tm\ll 1$ throughout the derivation, which is a valid assumption for our setup, in which $\Gup \tm=3\times 10^{-4}$. We can, therefore, assume that at most one excitation event occurs within a single measurement interval, reducing the number of possible events during a measurement interval to three:

\begin{enumerate}[leftmargin=*]
  \setcounter{enumi}{-1}

    \item No excitation occurs, and coherence is fully maintained, i.e. $C^{(0)}=1$. This event happens with probability $p_0 =  1-\Gup \tm$. The corresponding dephasing rate is zero: $\Gi{0}=0$. 

   \item The ancilla is excited, and remains in $\ket{e}$ until it is reset at $t=\tm$.
Since $\Gup \tm\ll 1$, the excitation probability is uniformly distributed over the interval. The probability density function for the excitation duration $T_1$ taking a value $\tau_1\in \left[0,\tm\right]$ is obtained by multiplying the probability density $\Gup$ of an excitation event at $\tm-\tau_1$ with the probability of no subsequent decay:
\begin{align*}
    f^{(1)}_{T_1}(\tau_1)&=\frac{1}{p_1}\Gup e^{-\Gq\tau_1},
\end{align*}
where we normalize by $p_1$, representing the total probability of this event: 

\begin{equation*}
p_1=\int_{\tau_1=0}^{\tm}{\Gup e^{-\Gq\tau_1} d\tau_1} = \Gup \tm \frac{1-e^{-\Gq\tm}}{\Gq\tm}.
\end{equation*}

For the experimentally relevant limit of a negligible ancilla decay rate, i.e., $\Gq\ll \tm^{-1},\chi$, we obtain that the event occurs with a probability $\tilde{p}_1=\Gup\tm$ and that the excitation duration is uniformly distributed with $\tilde{f}^{(1)}_{T_1}(\tau_1)=\frac{1}{\tm}$.
The corresponding cavity qubit coherence is given by Fourier transforming this uniform distribution, yielding

\begin{align*}
 C^{(1)} &= \langle e^{i\chi T_1}\rangle e^{i(\widetilde\theta+\theta_0)} \\
 &= e^{i(\widetilde\theta+\theta_0)}\int_{\tau_1=0}^{\tm} e^{i\chi\tau_1}\tilde{f}^{(1)}_{T_1}(\tau_1)\,d\tau_1 \\
 &= e^{i(\chi \tm/2+\theta_0+\widetilde\theta)}\cdot \mathrm{sinc}(\chi \tm/2),
\end{align*}
corresponding to \cref{eq:coherence_abs_and_phase}.
In this expression, we included the feedback phase $\widetilde\theta$ applied upon detection of the excitation, as well as a phenomenological offset phase $\theta_0$. Such an offset phase could result from a differential cavity Stark shift between the ground and excited states in the presence of a measurement drive. The dephasing rate due to this event is minimized by choosing an optimal feedback phase $\widetilde\theta_\mathrm{opt}$ such that $\bar\theta^{(1)}\equiv\arg(C^{(1)})=0$, yielding $\Gi{1} = \frac{1}{\tm} p_1\left(1- |C^{(1)}|\right)$. 
The optimal feedback phase in the limit of negligible ancilla decay is therefore  $\widetilde\theta_\mathrm{opt}=-\frac{\chi \tm} {2}\textrm{mod}\, \pi -\theta_0$. In the limit of $\Gq\ll \tm^{-1},\chi$, excitations without subsequent spontaneous decay (i.e., event 1) are the dominant contributors to dephasing, yielding a cavity pure dephasing rate 
 
\begin{equation}
\label{eq:feedback_high_frequency}
\Gphifb\approx\Gi{1}\approx\Gup\left(1-|\mathrm{sinc}(\chi \tm/2)|\right).
\end{equation}

This expression is used in \cref{eq:coherence_no_error_with_fb}. Had we not applied any phase correction, i.e., $\widetilde\theta = 0$, we would obtain instead $\Gphifb\approx\Gup(1-\mathrm{sinc}(\chi \tm))$, where we neglected the offset phase. For frequent measurements $\chi \tm \ll 1$, the phase correction therefore results in a fourfold reduction in decoherence.

The expression for the cavity qubit coherence without neglecting ancilla decay is given by
\begin{align*}
 C^{(1)}&=\langle e^{i\chi T_1}\rangle e^{i(\widetilde\theta+\theta_0)} \\
 &= e^{i(\widetilde\theta+\theta_0)}\int_{\tau_1=0}^{\tm} e^{i\chi\tau_1}f^{(1)}_{T_1}(\tau_1)\,d\tau_1 \\
 &=\frac{\Gup}{p_1}\frac{e^{(i\chi -\Gq)\tm}-1}{(i\chi-\Gq)}e^{i(\widetilde\theta+\theta_0)}.
\end{align*}
We note that since the probability distribution of the excitation duration  $f^{(1)}_{T_1}(\tau_1)$ is a multiplication of a uniform distribution and an exponential distribution, the coherence can be interpreted as a convolution of the characteristic functions of these distributions, i.e.  
$ C^{(1)} = e^{i(\widetilde\theta+\theta_0)}\frac{\Gq\tm}{1-e^{-\Gq\tm}}\left[e^{i\chi \tm/2}\mathrm{sinc}(\chi \tm/2)*\frac{1}{(i\chi-\Gq)}\right](\chi)$.
The absolute value of the cavity coherence and its phase
\begin{align}  
&|C^{(1)}|= \frac{\Gq}{1-e^{-\Gq\tm}} \cdot \frac{1}{\sqrt{\chi^2+\Gq^2}}\nonumber \\
&\quad\cdot \sqrt{[e^{-\Gq \tm}\cos(\chi \tm)-1]^2+[e^{-\Gq \tm}\sin(\chi \tm)]^2},
\label{eq:single_excitation_coherence_abs}
\end{align}
\begin{align}  
&\bar\theta^{(1)} = \nonumber\\
&\quad\tan^ {-1}  \left(\frac{\chi[e^{-\Gq \tm}\cos(\chi \tm)- 1]+\Gq e^{-\Gq \tm}\sin(\chi \tm) }{\Gq[e^{-\Gq \tm}\cos(\chi \tm)- 1]-\chi e^{-\Gq \tm}\sin(\chi \tm) } \right) \nonumber \\
&\quad+\widetilde\theta+\theta_0,
\label{eq:single_excitation_coherence_phase}
\end{align}
were used as the theoretical model for the data in \cref{fig:cavity_decoherence b}, which postselects on the occurrence of this event. 

    \item In the third and final event, an excitation occurs, but it decays before the ancilla is measured. This event happens with a probability $ p_2=1-p_0-p_1= \Gup \tm \left(1 - \frac{1-e^{-\Gq\tm}}{\Gq\tm}\right)$. The corresponding probability distribution $f^{(2)}_{T_1}(\tau_1)$ is obtained by summing over probabilities of events in which the excitation occurred at a time $s=[0,\tm-\tau_1]$ and lasted for a duration $\tau_1$: 
    
\begin{align*} 
    f^{(2)}_{T_1}(\tau_1)&= \frac{1}{p_2}\int_{s=0}^{\tm-\tau_1} \Gup e^{-\Gq \tau_1}\Gq \,ds \\
    &=\frac{1}{p_2}\Gup\Gq(\tm-\tau_1)e^{-\Gq\tau_1}.
\end{align*}

The resulting coherence is
    
    \begin{equation}
        C^{(2)} = \frac{\Gup \Gq }{p_2}\frac{\left[e^{(i\chi-\Gq)\tm}-1-(i\chi-\Gq)\tm\right]}{ (i\chi-\Gq)^2}.
    \end{equation}
 
No feedback is applied for this event, as the ancilla decays before the excitation can be observed. In the limit of infrequent measurements $\tm^{-1}\ll \Gq,\chi$, this event dominates the dephasing rate, and we retrieve the no-feedback pure dephasing rate of \cref{eq:gamma_phi_cav_no_fb} in the main text:
\begin{align}
 \Gphifb\approx\Gi{2}\approx\Gup\frac{\chi^2}{\chi^2+\Gq^2}.
 \label{eq:coherence_rate_infrequent_measure_rate}
\end{align}

\begin{figure*}[t!]
    \centering
    \includegraphics[scale=0.37]{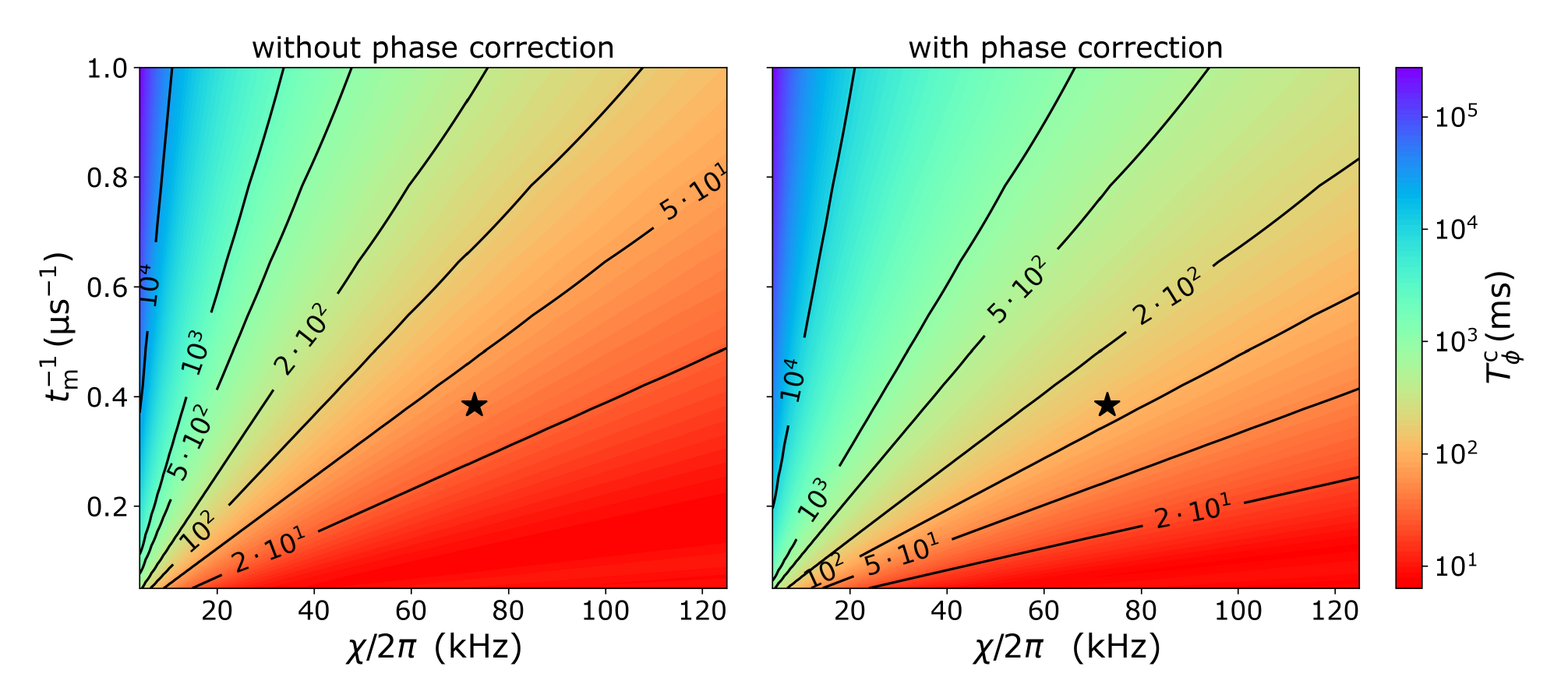}
    \caption{\label{fig:supp_phase_correction_theory} \textbf{Pure dephasing time with and without phase correction}. Cavity qubit dephasing time versus the dispersive interaction rate $\chi$ and the ancilla measurement rate $\tm^{-1}$ without (left panel) and with (right panel) cavity phase correction. The stars denote the dephasing times for the parameters from this work, namely, $\chi/2\pi= 73.06\,$ MHz and $\tm^{-1}=(2.6\, \mathrm{\mu s})^{-1}$.}
    
\end{figure*}

\end{enumerate}

\subsection{Performance of the ideal coherence recovery protocol}

In \cref{fig:supp_phase_correction_theory}, we show the predicted pure dephasing time $\Tphifb=1/\Gphifb$ as a function of $\chi$ and $\tm$, comparing the performance of the optimal phase feedback to that without phase feedback. With the parameters of our experimental system, we obtain dephasing times of 122 ms and 35 ms with and without phase feedback, respectively. The dephasing time in both cases is mostly limited by events of type 1, which account for 96\% of all excitation events. When postselecting on zero detections, i.e., considering only events of type 2, the dephasing time increases to 1.7 s. In practice, the measured postselected dephasing time is lower due to measurement-induced shot-noise dephasing (see \cref{sec:measurement_errors}). It is also apparent that a modest increase in the measurement rate can yield pure dephasing times close to a second without compromising on the dispersive interaction strength.


\section{effect of non-ideal measurements on  cavity decoherence}
\label{sec:measurement_errors}

\begin{figure}[b!]
    \centering
    \includegraphics[scale=0.37]{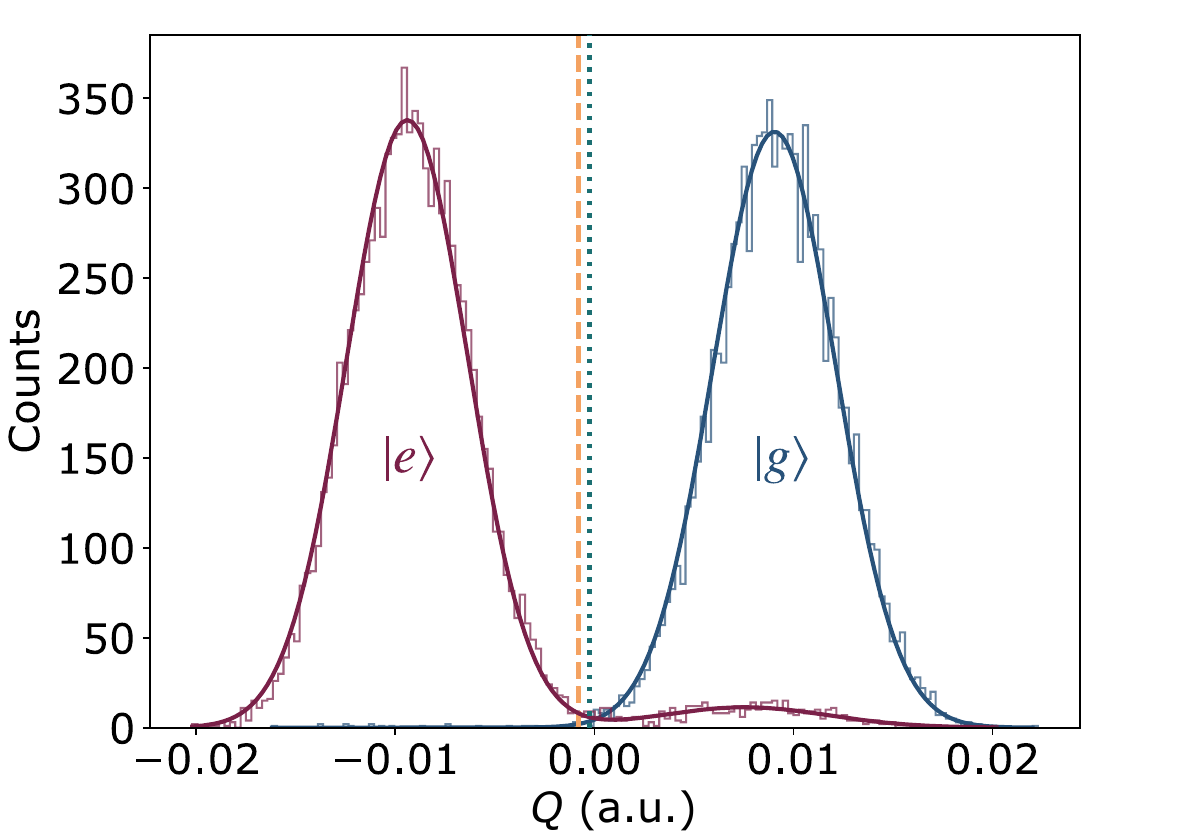}
    \caption{\label{fig:supp_readout_histograms} \textbf{Single-shot ancilla measurement}. Histogram of the integrated quadrature $Q$ of a single-shot measurement after preparing the transmon ancilla in either the ground state (blue) or excited state (red). The dashed and dotted lines mark the decision boundaries $\eta$ for assigning the measurement outcome $\ket{e}$ (left of the boundary) or $\ket{g}$ (right of the boundary). The decision boundary can be set to optimize the discrimination fidelity (dotted line) or to optimize the performance of the coherence recovery protocol (dashed line). The latter results in a bias towards lower false-positive probabilities. The low-amplitude Gaussian to the right of the boundary, following ancilla initialization in $\ket{e}$, arises from decay events occurring during the measurement. Within the coherence recovery protocol, this event corresponds to a decayed ancilla excitation (event 2), rather than a measurement error.}
    
\end{figure}

\subsection{Description of the ancilla measurement}
In this section, we describe the experimental implementation of the repeated transmon ancilla measurements, which are crucial for achieving a significant gain with our protocol.
Each measurement involves sending a pulse to the readout resonator and analyzing the reflected signal, which is then integrated into a single quadrature $Q$. This quadrature is used to discriminate between two hypotheses: whether the ancilla is in the ground state or the excited state:
\begin{align*} Q &> \eta \,: \quad \text{ancilla in $\ket{g}$} \\ Q &\le \eta \,:\quad \text{ancilla in $\ket{e}$}, \end{align*}
where $\eta$ is the decision boundary. 
The amplitude and duration of the readout pulse were optimized to balance two competing factors: While increasing them improves state discrimination, it also raises the photon population in the readout resonator, leading to increased measurement-induced shot-noise dephasing of the cavity qubit. In essence, measurement-induced shot-noise dephasing arises from the nonzero cross-Kerr interaction $\chi_{\text{cr}}$ between the readout resonator and the cavity, causing the readout pulse to weakly measure the cavity state. The readout drive frequency was placed symmetrically in between the two state-dependent readout resonator frequencies. This choice of drive frequency suppresses cavity shot-noise dephasing due to a relatively low-readout photon-number population, while maintaining a high measurement fidelity.

\cref{fig:supp_readout_histograms} shows histograms of the integrated quadrature $Q$ after initializing the transmon ancilla in either $\ket{g}$ or $\ket{e}$, demonstrating the discrimination performance using the optimized pulse with a duration of $t_\mathrm{RO} = 1.6$ \mus. 
The discrimination fidelity $\mathcal{F}_d $ represents the probability of correctly identifying the \textit{projected} state and is given by \cite{varbanov2020leakage}
\begin{equation}
    \mathcal{F}_d = 1 - \frac{p_{e|g}+p_{g|e}}{2},
\end{equation}
where $p_{e|g}$ is the probability of a false positive (misidentifying an ancilla in $\ket{g}$ as being in $\ket{e}$) and  $p_{g|e}$ is the probability of a false negative (misidentifying an ancilla in $\ket{e}$ as being in $\ket{g}$). Both errors arise from the finite signal-to-noise ratio of the measurement. The discrimination fidelity differs from the more commonly used assignment fidelity, which corresponds to the probability of correctly identifying the $\textit{initial}$ state \cite{chen2023transmon}. Unlike the discrimination fidelity ($\mathcal{F}_d=99.87\%$ in our case), the assignment fidelity ($\mathcal{F}_a=97.7\%$) accounts for ancilla relaxation during the measurement, treating it as a measurement error. However, in our protocol, relaxation of an excited ancilla during the measurement is considered a desired outcome if the state is identified as $\ket{g}$. This occurrence is captured by events of type 2, representing decayed ancilla excitations (see \cref{sec:coherence_derivation}). Therefore, discrimination fidelity is the relevant metric for evaluating measurement performance. 
 
 The discrimination fidelity is maximized by setting the decision boundary $\eta$ at the intersection of the two Gaussian quadrature distributions (see \cref{fig:supp_readout_histograms}), where $p_{g|e}=p_{e|g}=0.13\%$, yielding $\mathcal{F}_d = 99.87\%$. However, this boundary $\eta$ is not optimal for our protocol, since the ancilla predominantly remains in $\ket{g}$. As a result, false positives are far more detrimental than false negatives.
We perform detailed simulations to find the decision boundary that maximizes the pure dephasing time of the cavity qubit. To ensure that the simulation accounts for the effects of repeated measurements, we used a hidden Markov model (HMM) to estimate the measurement and ancilla parameters based on the full history of measurement outcomes (see \cref{sec:markov}). This optimization leads to a boundary slightly biased to the left of the intersection between the two Gaussian quadrature distributions, resulting in a reduced false-positive probability of $p_{e|g} = 2.16\times10^{-4}$ and an increased false-negative probability of $p_{g|e} = 1.4\times 10^{-2}$. 

Finally, we adopted a readout pulse shape from Refs. \cite{jerger2024dispersivequbitreadoutintrinsic,motzoi2018simple}, which ensures that the readout resonator returns to vacuum after each measurement, irrespective of the ancilla state. This process allows us to apply the readout pulses at a high repetition rate (up to $(2.3$ \mus$)^{-1}$), while preventing each measurement from impacting subsequent ones. 

In future work, a real-time state estimator based on a HMM could be implemented to account for the full measurement history during the coherence recovery protocol. Additionally, the decision boundary could be adapted in real time. These methods have the potential to more accurately determine the state of the system and hence further improve the performance of the protocol.

\subsection{Simulated error budget of the coherence recovery protocol in the presence of nonideal measurements}

In this section, we modify the coherence expressions to account for nonideal measurements.
The first nonideality we include is measurement-induced dephasing due to photon shot noise in the readout resonator. We denote the measurement-induced cavity decoherence from a single measurement as $C_\mathrm{RO}$. 

The second non-ideality we address is imperfect measurement fidelity, which results in false-positive and false-negative events. 
When the ancilla state is incorrectly identified, the reset pulse and phase feedback are applied when they should not be and vice versa. As a result, the ancilla will spend the subsequent interval $\tm$ in the excited state, assuming no further errors occur. This finding leads to a deterministic phase accumulation $e^{i\chi\tm}$ in the subsequent interval, which should be taken into account when deriving the coherence. While the effect of measurement errors on the subsequent interval violates the independence assumption used to derive the cavity dephasing rate, the error introduced by this assumption is only a small second-order effect and can be ignored. 

\begin{figure}[b!]
    \centering
    \includegraphics[scale=0.63]{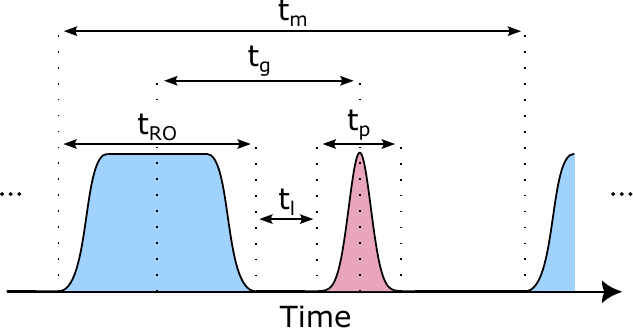}
    \caption{\label{fig:supp_pulse_sequence} \textbf{Timing diagram of the coherence recovery protocol}.  Measurement pulses, shown in blue, have a duration of $t_\mathrm{RO}=$1.6 \mus, with transmon ancilla state inference occurring on average at the pulse's midpoint. The latency ($t_\textrm{l}=$372 ns) between the end of the measurement pulse and the start of the conditional reset pulse ($t_\mathrm{p}=$128 ns) is attributed to processing and signal propagation delays. The gap time,  $t_\textrm{g}=\frac{t_\mathrm{RO}}{2}+t_\textrm{l}+\frac{t_\mathrm{p}}{2}$, marks the duration between the transmon ancilla state inference and the midpoint of the reset pulse. The time between measurements  $t_\textrm{m}\ge t_\mathrm{RO}+t_\textrm{l}+t_\textrm{p}$, is independently set for each experiment.}
    
\end{figure}

Finally, we consider a gap $t_\mathrm{g}=1.24\,\mathrm{\mu s}$ between the detection time, which we assume occurs halfway between the measurement pulse and the reset pulse (see \cref{fig:supp_pulse_sequence}). This gap encapsulates the combined effects of the finite readout duration, the feedback latency ($t_\mathrm{l}=372$ ns), and the reset pulse duration ($t_\mathrm{p}=$128 ns). If the ancilla is excited at the time of the detection, the gap results in a deterministic cavity phase accumulation $e^{i\chi t_\mathrm{g}}$. Additionally, the ancilla may decay during this gap period, causing the reset pulse to reexcite the ancilla.

We now enumerate the possible events using the same numbered notation as in the previous section, with letters added to indicate specific measurement scenarios.

\begin{enumerate}[leftmargin=*]
  \setcounter{enumi}{-1}
  
\item (a) No excitation occurs, and the state is correctly detected as $\ket{g}$. This event is the most likely one, with probability $p_{0a} =  (1-\Gup \tm) p_{g|g}$. The coherence is affected due to readout-induced shot-noise dephasing, yielding $C^{(0a)}=C_\mathrm{RO}$.

(b) No excitation occurs, but the state is incorrectly identified as $\ket{e}$. The associated probability is $p_{0b} =  (1-\Gup \tm)p_{e|g}$. The reset pulse excites the ancilla to $\ket{e}$, where it remains for a duration $\tm$ assuming no decay occurs. The subsequent measurement correctly identifies the state as $\ket{e}$, and the ancilla is reset to $\ket{g}$. Hence, the feedback phase is applied twice, while the offset phase is applied only once. The coherence is therefore $C^{(0b)}=C_\mathrm{RO}e^{i(2\widetilde\theta+\chi\tm+\theta_0)}$.
Such false-positive events result in an effective heating of the ancilla at a rate $\frac{p_{e|g}}{\tm}$, acting as an independent cavity dephasing mechanism to thermal excitations.
However, since the phase feedback for the dominant event (1a, see below) is approximately $\widetilde\theta \approx -(\frac{\chi \tm}{2}+\theta_0+\chi t_\mathrm{g})$ , the phase $\chi \tm+\theta_0$ acquired per false positive is partially mitigated by the application of the twofold phase feedback $2\widetilde\theta$. 

\item (a) The ancilla is excited and does not decay. The excitation is correctly identified during the measurement and does not decay during the gap period. The associated probability and coherence are $p_{1a} =  p_1 p_{e|e} e^{-\Gq t_\mathrm{g}}$ and
$C^{(1a)}=C_\mathrm{RO} C^{(1)} e^{i\chi t_\mathrm{g}}$, where $p_1$,$C^{(1)}$ are the probability and coherence derived for this event with ideal measurement in the previous section, and $\chi t_\mathrm{g}$ is the phase acquired during the gap period before the ancilla is reset to $\ket{g}$.

 (b) The ancilla is excited and does not decay, but the ancilla state is incorrectly identified as $\ket{g}$. This event occurs with a probability $p_{1b} =  p_1 p_{g|e} e^{-\Gq t_\mathrm{g}}$. The phase feedback is applied after the subsequent interval, assuming the ancilla does not decay and is correctly identified as $\ket{e}$.
 The coherence is therefore $C^{(1b)}=C_\mathrm{RO} C^{(1)} e^{i(\chi(\tm+t_\mathrm{g})+\theta_0)}$.

 (c) The ancilla is excited and does not decay. The excitation is correctly identified, but it decays during the gap period. On average, the ancilla spends another period $t_\mathrm{g}/2$ in its excited state, decays, and is then reexcited by the reset pulse. The probability of this event is
  $p_{1c} =  p_1 p_{e|e} (1-e^{-\Gq t_\mathrm{g}})$. The feedback and offset phases are applied twice and the associated coherence is
  $C^{(1c)}=C_\mathrm{RO} C^{(1)} e^{i(\widetilde\theta + \chi(\tm+t_\mathrm{g}/2)+\theta_0)}$.
  
\item (a) An excitation occurs, but it decays before the ancilla is measured and correctly identified as $\ket{g}$. This event happens with a probability $p_{2a} = p_2 p_{g|g}$. The corresponding coherence is  $C^{(2a)}=C_\mathrm{RO}C^{(2)}$. 

(b) An excitation occurs, but it decays before the ancilla is measured. The state is incorrectly identified as $\ket{e}$, resulting in the application of two feedback phases and a single offset phase. The probability is $p_{2b} = p_2 p_{e|g}$ and the coherence is 
$C^{(2b)}=C_\mathrm{RO} C^{(2)} e^{i(2\widetilde\theta+\chi\tm+\theta_0)}$. The probability of this event is negligible, but we mention it for completeness.

\end{enumerate}

\begin{figure}[h!]
    \centering
    \includegraphics[scale=0.35]{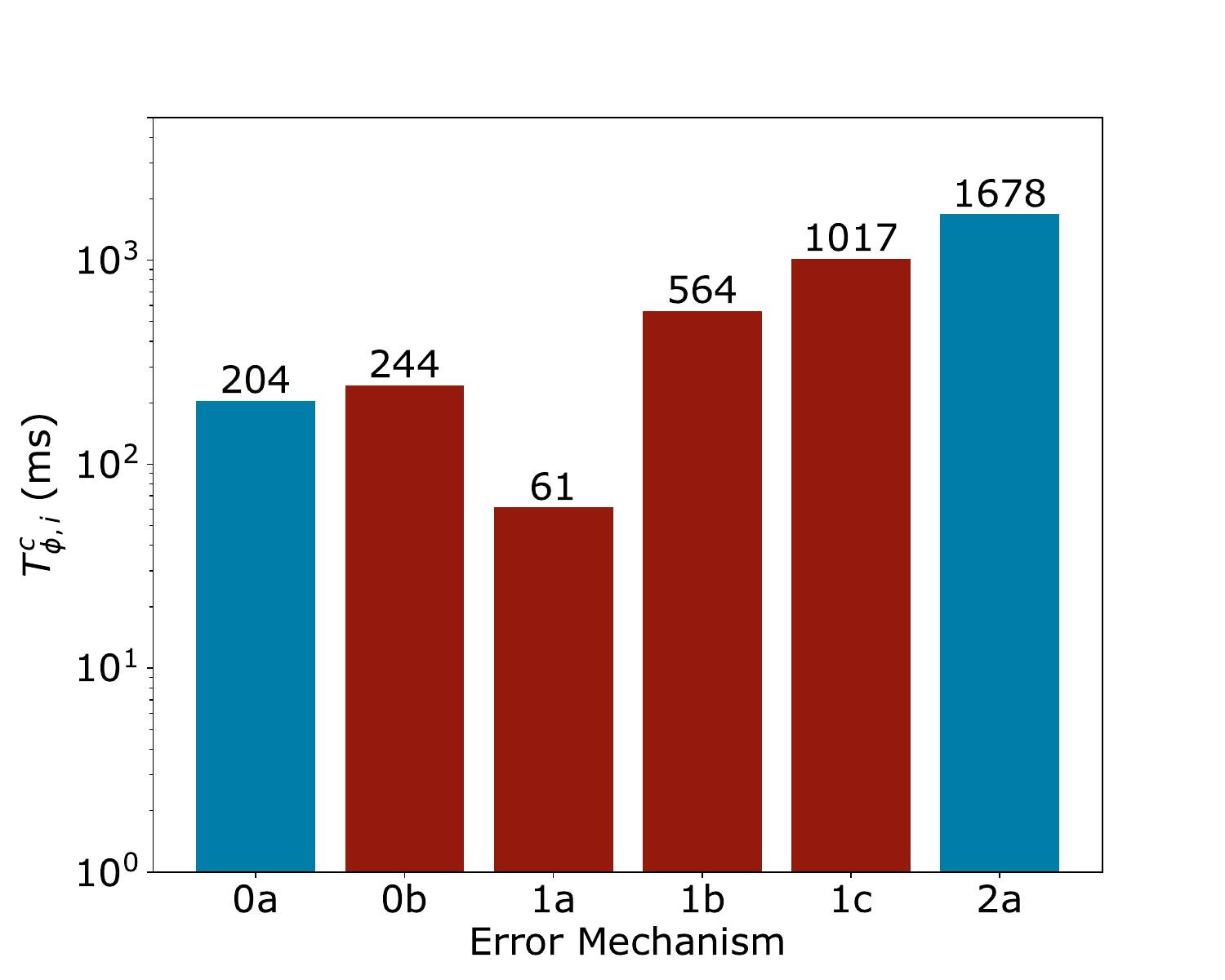}
    
    \caption{\label{fig:supp_error_mechanism} \textbf{Error budget of the coherence recovery protocol}. The cavity pure dephasing time associated with each mechanism $T^\mathrm{c}_{\phi,i}=1/\Gi{i}$ is shown according to the list in  \cref{sec:measurement_errors}. The calculations are based on the experimental parameters provided in \cref{table:1}, with $\tm=2.6$ \mus\, and the optimal phase determined using \cref{eq:optphase}.  
    Event classes in blue contribute to the ``erasure'' dephasing time, i.e., those remaining after postselecting on no observed excitations. Error mechanism 1b is excluded from this set, as it is most likely to be detected in the subsequent measurement interval. The value for error mechanism 0a, representing the readout-induced shot-noise dephasing $C_\mathrm{RO}$, was obtained by subtracting the calculated contribution of event 2a, $\Gamma_{\phi, 2a}^c=(1678\,\mathrm{ms})^{-1}$, from the observed erasure dephasing rate of $(182\,\mathrm{ms})^{-1}$.   
   The total lifetime predicted from this error budget is $T^\mathrm{c}_{\phi}=1/\sum\Gi{i}= 34.8\, \mathrm{ms}$, in agreement with the measured value of \vunit{35}{4}{ms}.}
\end{figure}

As in the previous section, the total cavity dephasing is given by
 \begin{equation}
 \label{eq:full_dephsing_theory}
    \Gphifb = \frac{1}{\tm}\sum_i {p_i\left(1- |C^{(i)}|\cos(\bar\theta^{(i)})\right)},
\end{equation}
where $\bar\theta^{(i)}=\arg(C^{(i)})$. The optimal feedback phase now has to strike a balance between the various events described above. Writing $\bar\theta^{(i)}=\bar\theta^{(i)}_0+\alpha_i \widetilde\theta$, with $\bar\theta^{(i)}_0$ the zero-feedback coherence phase of the $i^\mathrm{th}$ event and $\alpha_i $ the prefactor of the feedback phase for that event as derived above, we can find the optimal angle by solving for $\partial_{\widetilde\theta} \Gphifb =0$. This procedure yields
 \begin{equation}
    \widetilde\theta_\mathrm{opt} = -\frac{ \sum_i {p_i |C^{(i)}|\alpha_i\bar\theta^{(i)}_0} }{\sum_i {p_i |C^{(i)}|\alpha_i^2} },
    \label{eq:optphase}
\end{equation}
where we used the approximation $\sin(\bar\theta^{(i)})\sim \bar\theta^{(i)}$ , which is valid for short readout intervals.
In \cref{fig:supp_error_mechanism}, we show the contribution of each dephasing mechanism when using the optimal phase feedback.

\section{Uncertainty in decay time}

\begin{figure}[h!]
    \centering
    \includegraphics[scale=0.45]{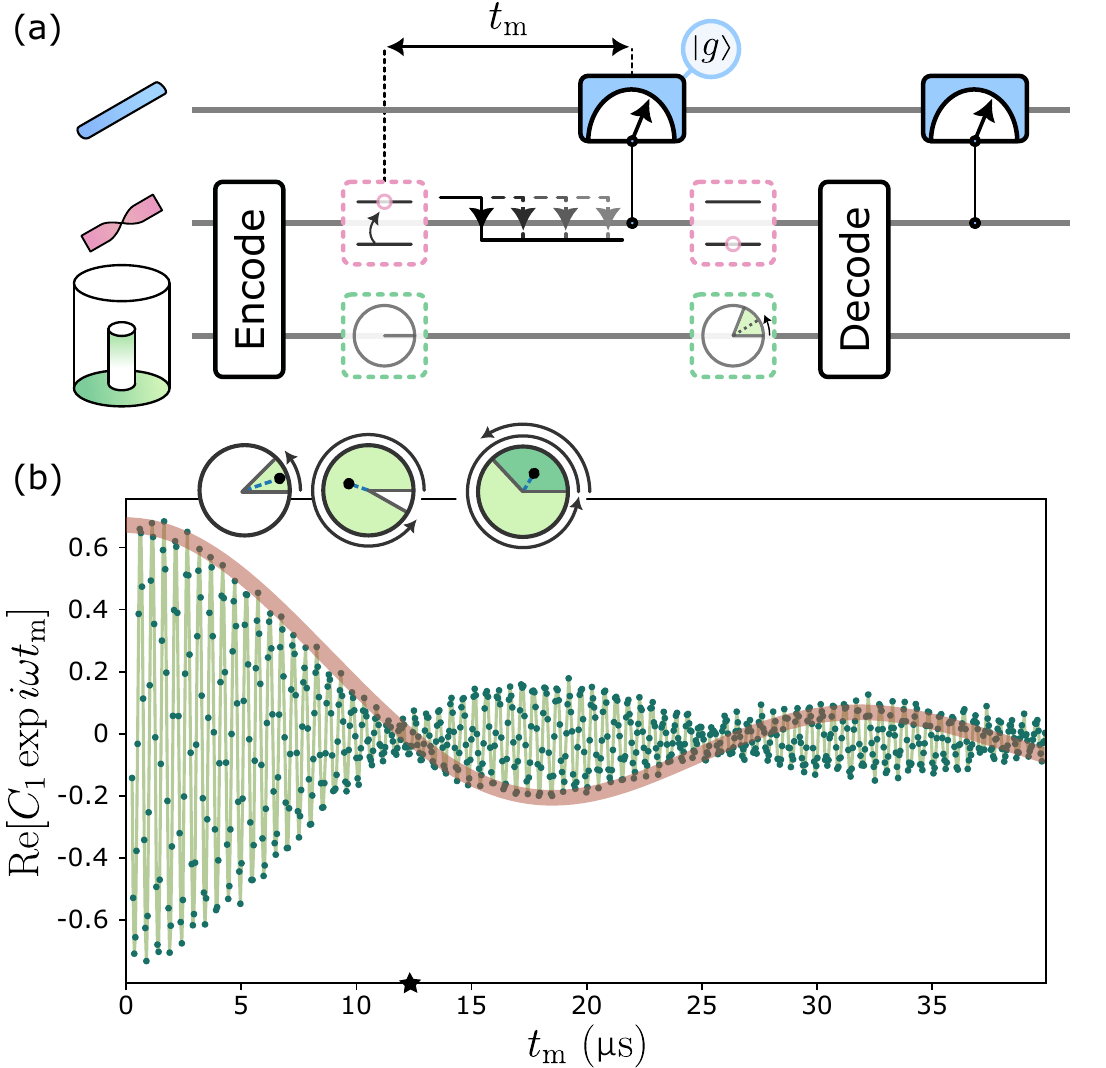}
    \phantomsubfloat{\label{fig:supp_ancilla_error a}}
	\phantomsubfloat{\label{fig:supp_ancilla_error b}}
    \vspace{-1\baselineskip}
    \caption{\label{fig:supp_ancilla_error} \textbf{Cavity decoherence due to ancilla relaxation.} \textbf{a}, Circuit for observing cavity decoherence from a single transmon ancilla relaxation event. The ancilla is initialized in its excited state, and the cavity is encoded in a superposition state $\frac{1}{\sqrt{2}}(\ket{0}+\ket{1})$. After a time $\tm$, we measure the ancilla and decode the cavity state postselecting on the observation of the ancilla in its ground state. \textbf{b}, Real part of the coherence of the cavity qubit as a function of the measurement time $\tm$ (green markers). For clarity, we use a frame in which the cavity rotates at a frequency $\omega = $\vunit{1.950}{0.95}{MHz}. The measurement interval corresponding to $\tm = 2\pi/\chi$ is indicated by a black star. The insets illustrate the equatorial plane of the cavity qubit’s Bloch sphere, with the phase distribution in green and the coherence indicated by a black dot. The green line corresponds to the real part of the coherence from \cref{eq:coherence_one} using the experimental parameters from \cref{fig:cavity_decoherence}, after multiplying by $e^{i\omega \tm}$ and rescaling to account for the imperfect preparation and measurement fidelities. The brown curve shows the real part of the coherence in a nonrotating frame.}    
\end{figure}

In \cref{fig:cavity_decoherence}, we explored the decoherence due to a single excitation after initializing the transmon ancilla in $\ket{g}$. In this appendix, we consider the opposite scenario: We prepare the ancilla in $\ket{e}$ and postselect on observing  $\ket{g}$, leaving only those experimental runs in which a relaxation event occurred (see \cref{fig:supp_ancilla_error a}). \cref{fig:supp_ancilla_error b} shows the resulting cavity coherence as a function of $\tm$. Since the decay rate is much higher than the excitation rate, this experiment has a much better signal-to-noise ratio than its counterpart in \cref{fig:cavity_decoherence b}.

The probability density for the excitation duration in this scenario is given by 
\begin{align*}
    f_T(\tau)&=\frac{\Gq e^{-\Gq\tau}}{1 -  e^{-\Gq \tm}} ,
\end{align*}
and the corresponding cavity coherence is
\begin{equation}
 C_1 = \int_{\tau=0}^{\tm} e^{i\chi\tau}f_T(\tau)\,d\tau=\frac{\Gq}{1 -  e^{-\Gq \tm}}\frac{ e^{(i\chi-\Gq)\tm}-1}{i\chi-\Gq}.
 \label{eq:coherence_one}
\end{equation}
This coherence is in excellent agreement with the data in \cref{fig:supp_ancilla_error b}.

\section{Hidden Markov model}
\label{sec:markov}
Repeated measurements influence the transmon ancilla parameters. For example, they can increase its temperature or affect its decay rate. Additionally, the readout performance may differ from that of independent single-shot measurements. In this section, we describe how we analyze these effects and characterize the measurement errors. Thus, we use a HMM. We introduce the HMM framework \cite{baum1966statistical}, explain how it can be used for parameter estimation \cite{dixit2021searching}, and discuss its role in optimizing the protocol and analyzing sources of error.

\subsection{Statistical model-free evolution}
We model the time-dependent evolution of a two-level transmon ancilla using a Markov process, which assumes the system's future state depends only on its current state, and not on the prior sequence of states. This Markov model is extended to a hidden Markov model due to the presence of measurement errors, meaning that the true ancilla state is hidden and must be inferred from noisy measurements. We denote the hidden ancilla state at time $n$ as $x_n$, and the corresponding measurement outcome as $o_n$. We use a colon to describe a collection of measurement outcomes, e.g., $o_{1:n} \equiv \{o_1,\,o_2,\, \ldots,\,o_n\}$. The hidden state transition probabilities are captured by a \textbf{transition} matrix, while the likelihood of measurement outcomes given a hidden state is described by an \textbf{emission} matrix:

\begin{enumerate}
    \item \textbf{Transition matrix}: determines the likelihood of the hidden state at the $n^\mathrm{th}$ time step, based on the hidden state's probability at the previous time step:   
    \begin{align*}
 T_{ij} \equiv \pr{x_n=j}{x_{n-1}=i}.
\end{align*}
We express the transition matrix using the probabilities for excitation $p_{\uparrow}=\Gup \tm$ and relaxation $p_{\downarrow}=1-e^{-\Gq \tm}$:
\begin{align*}
T = \begin{pmatrix}
	1 - p_{\uparrow} & p_{\uparrow} \\ 
	p_{\downarrow} &  1 - p_{\downarrow} \\
\end{pmatrix}
\end{align*}

\item \textbf{Emission matrix}:  
accounts for the imperfections in our measurement process. It describes the probability of observing a measurement outcome given the actual (hidden) state of the ancilla. The elements of the emission matrix $M_{jk}$ represent the probability of observing outcome $k$ when the true state is $j$:
\begin{align*}
    M_{jk} \equiv \pr{o_n=k}{x_n=j}.
\end{align*}
We can write it in a matrix form using the discrimination errors:
\begin{align*}
M = \begin{pmatrix}
	1 - p_{e|g} & p_{e|g}\\
	p_{g|e} & 1 -p_{g|e}
\end{pmatrix},
\end{align*}
where $p_{e|g}$ and $p_{g|e}$ are the probabilities for false-positive and false-negative observations. To incorporate this matrix in the calculation we use an indicator column vector to describe the observations: 
\begin{align*}
    \gamma(o_i) \equiv \begin{cases}
        (1,\, 0)^T & \text{if }o_i = 0 \\
        (0,\, 1)^T & \text{if }o_i = 1 .\\
    \end{cases}
\end{align*} 
Using the indicator, we can express the probability of obtaining the outcome $o_n$ given the hidden state $x_n=j$:
\begin{align*}
	\pr{o_n}{x_n=j} = \sum_{k}M_{jk}\cdot \gamma(o_n)_k.
\end{align*}

\item \textbf{Initial vector}: this vector, denoted by $\pi$, describes the probability of the hidden state at the onset of the evolution. 

\end{enumerate}

\subsection{Inference - likelihood and smoothed values}
\label{sec:estimation_of_parameters}
Next, we derive a statistical inference model to estimate the HMM parameters. 
Specifically, we aim to reconstruct the matrices $T, M$ by finding the set of parameters $\lambda=\{p_{e|g},\, p_{g|e}, \,\Gup, \,\Gq\}$ that best explain the observed data. Formally, we want to maximize the posterior distribution. The optimal estimate $\bar{\lambda}$ satisfies
\begin{align*}
    \bar{\lambda} &= \argmax_{\lambda}\Big[p(\lambda|o_{1:N})\Big] \\
	&= \argmax_{\lambda}\Big[\mathcal{L}_\lambda(o_{1:N})\cdot p(\lambda)\Big]
 \approx \argmax_{\lambda}\Big[\mathcal{L}_\lambda(o_{1:N})\Big],
\end{align*}
 where the likelihood $\mathcal{L}_{\lambda}$ that $\lambda$ matches the observations $(o_{1:N})$ and $N$ is the final time step. 
We can derive the likelihood by establishing a recursive relation by either going forward (from start to end) or backward (from end to start):
\begin{enumerate}[leftmargin=*]
    \item \textbf{forward} -  
    \begin{align*}
        \mathcal{L}_{\lambda}(o_{1:N}) \equiv p(o_{1:N}|\lambda)  = \sum_{x_N}p(o_{1:N}, x_N\,|\lambda) \equiv \sum_{i}f_i(N),
    \end{align*}    
    where the forward vector $f_i(k)$ is the joint probability of the hidden state at time $k$ being $i$ and the observations up to time $k$. We can now expand the forward vector recursively using the emission and transition matrices:
    \begin{align*}
	f_i(k) = \sum_{l,j}\gamma(o_k)_l\cdot M_{il}\cdot T_{ji}\cdot f_j(k-1),
\end{align*}
where the initial condition is $f_i(0) = \pi_i$. Next, we introduce a forward vector that is normalized at each time step, denoted by $\hat{f}_i(k)\equiv \frac{{f}_i(k)}{{f}_0(k)+{f}_1(k)}$.
We can express the normalized forward vector recursively:
\begin{align*}
\hat{f}_i(k) =  c_k^{-1}\sum_{l,j}\gamma(o_k)_l\cdot M_{il}\cdot T_{ji}\cdot \hat{f}_j(k-1) = \frac{f_i(k)}{\prod_{s=1}^{k}c_s},
\end{align*}
where $c_k$ is a normalization scalar derived from the recursive relation:
\begin{align*}
    c_k = \sum_i\sum_{l,j}\gamma(o_k)_l\cdot M_{il}\cdot T_{ji}\cdot \hat{f}_j(k-1).
\end{align*}
\item \textbf{backward} - 
\begin{align*}
    \mathcal{L}_{\lambda}(o_{1:N}) &\equiv p(o_{1:N}|\lambda) = \sum_{i}\pr{o_{1:N}}{x_0=i, \lambda}\cdot \pr{x_0 = i}{\lambda} \\
    &\equiv 
    \sum_{i}b_i(0)\cdot \pi_i,
\end{align*}
where the backward vector $b_i(k)$ is the probability vector of the observations from time $k+1$ until the final time step, given that the hidden state at time $k$ is equal to $i$:
\begin{align*}
	b_i(k) &= \pr{o_{k+1:N}}{x_k=i, \lambda} \\
 &= \sum_{j, l}\gamma(o_{k+1})_l\cdot M_{jl}\cdot T_{ij} \cdot b_j(k+1),
\end{align*}
where the boundary condition is $b_i(N) = (1,\, 1)$. Similarly to the forward vector, we can normalize the backward vector at each time step, denoting the normalization coefficient as $c_{k+1}$. We then obtain for each time step $k$:
\begin{align*}
\hat{b}_i(k) = c_{k+1}^{-1}\sum_{j, l}\gamma(o_{k+1})_l\cdot M_{jl}\cdot T_{ij} \cdot \hat{b}_j(k+1) = \frac{b_i(k)}{\prod_{s=k+1}^Nc_s}.
\end{align*}
\end{enumerate}

Finally, we can combine both forward and backward vectors to compute a ``smoothed value'' for the hidden variable at time step $k$. Formally, we denote $\omega_i(k)$ as the probability for $x_k=i$ given all available observations:
\begin{align*}
	\omega_{i}(k) &= \pr{x_k=i}{o_{1:N}, \pi} = \frac{\pr{x_k=i, o_{1:N}}{\pi}}{\pr{o_{1:N}}{\pi}} \\
	&=
	\frac{b_i(k)\cdot f_i(k)}{\sum_{i}b_i(k)\cdot f_i(k)} 
	= \hat{b}_i(k)\cdot \hat{f}_i(k).	
\end{align*}
Using the smoothed values we can estimate the most probable hidden states of the ancilla given a set of system parameters and observations. Conversely, the smoothed values can be used to estimate the system parameters by employing an optimization algorithm, In essence, such an algorithm estimates the hidden state given a set of observations, updates the system parameters accordingly, and repeats this process.

\subsection{Parameter estimation and reconstruction}

\begin{figure}[h!]
    \centering    
    \includegraphics[scale=0.4]{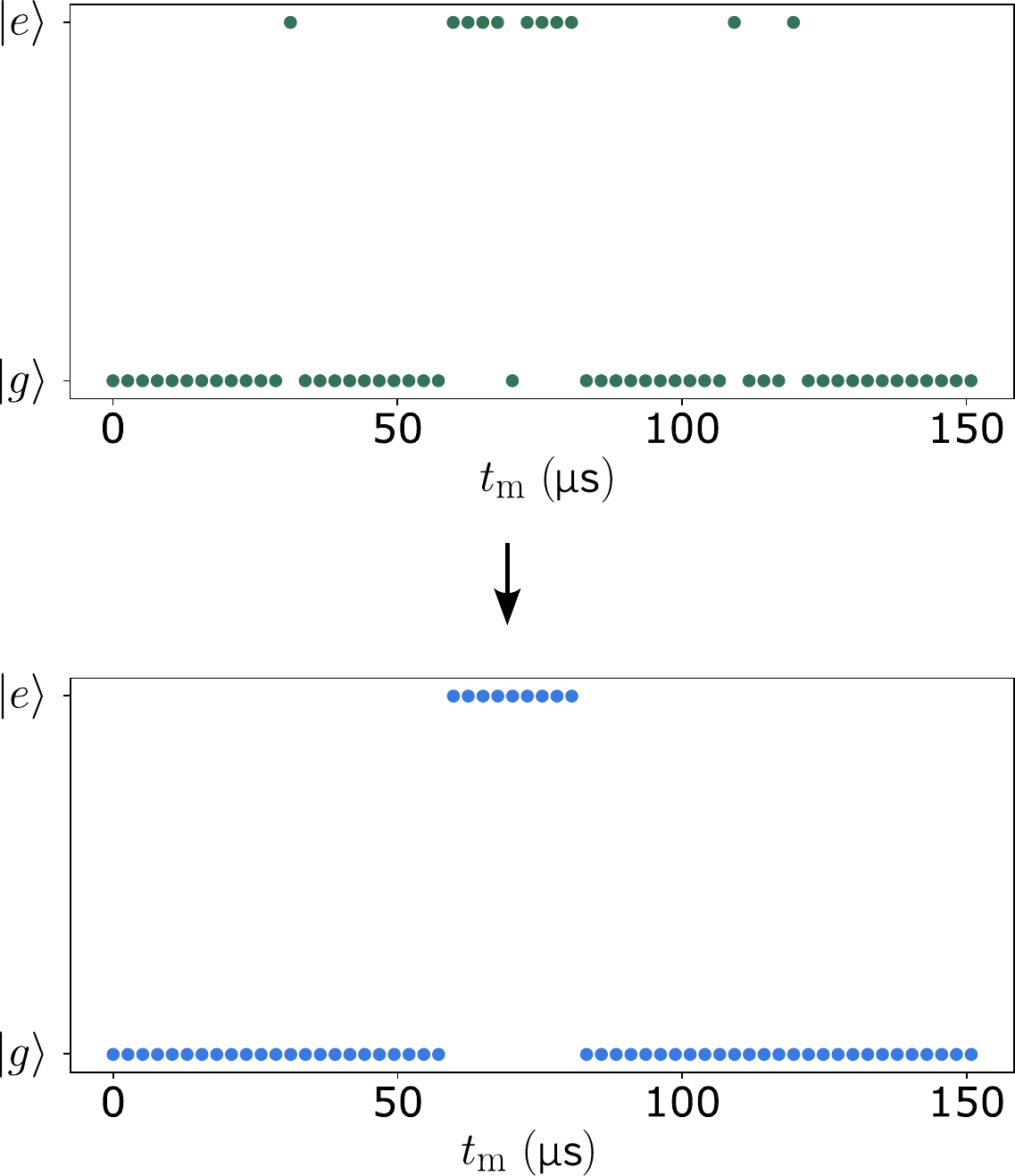}
    \caption{\label{fig:supp_markov} \textbf{Markov estimation and reconstruction procedure}. The upper panel shows a simulated set of observations. The lower panel demonstrates the reconstructed state after estimating the parameters from the observations. In particular, the reconstruction successfully identifies three false-positive events, a single ancilla excitation, and a single false-negative event. The time between measurements is $\tm=2.6\, \mathrm{\mu s}$. The parameters estimated from the simulated observations are: $p_{e|g} = 6\%$, $p_{g|e} = 14\%$, $\Gq = 56$ kHz, and $\Gup = 10$ kHz.}    
\end{figure}

We choose the Baum-Welch algorithm \cite{baum1970maximization} for finding the parameters that best explain the observations. The Baum-Welch algorithm is an expectation-maximization algorithm suitable for HMM that uses both forward and backward vectors. The parameter estimation process starts by recording the trajectory of the transmon ancilla using repeated measurements. We then feed the optimizer the observed data, obtaining the most likely model parameters $\{p_{e|g},\, p_{g|e},\, \Gup,\, \Gq\}$. We show an example of this process in \cref{fig:supp_markov}, where we use simulated data. Using the estimated parameters, we can next compute the smoothed values, yielding the most likely trajectory of the ancilla state.  

\subsection{Application of HMM in our work}
We use HMM-based parameter extraction to characterize the effects of repeated measurements on discrimination errors, ancilla relaxation, and ancilla temperature. For example, we observe a doubling of the ancilla relaxation rate when performing repeated measurements. These parameters were crucial for accurately modeling the cavity decoherence rates in our protocol, as detailed in the error budget section, and for ensuring the reliability of our simulations. Additionally, this method was instrumental in calibrating the ancilla heating drive in \cref{fig:feedback_perf a}. We used the parameter extraction to map the drive amplitude to the ancilla excitation rate, allowing us to identify the amplitude that would yield the desired value.

The HMM could also be used to implement a real-time estimator of the ancilla state, accounting for the full measurement history. This approach could mitigate the impact of imperfect measurement fidelities on the coherence recovery protocol. The implementation of this improved protocol is left for future work.

\end{document}